\documentclass[%
aps,
twocolumn,
superscriptaddress,
nofootinbib,
 amsmath,amssymb,
prd,
nopacs,
longbibliography
]{revtex4-2}

\usepackage[colorlinks,linkcolor=blue,citecolor=blue,urlcolor=blue,hyperindex,driverfallback=dvipdfm]{hyperref}

\usepackage{upgreek}
\usepackage{amsmath,amssymb,amsfonts}
\usepackage{appendix}
\usepackage{bm}
\usepackage{color}
\usepackage[normalem]{ulem} 

\usepackage{textcomp}

\usepackage{graphicx}
\usepackage{indentfirst}
\usepackage{psfrag}
\usepackage{epsfig}

\usepackage{svg}
\DeclareMathOperator{\sech}{sech}

\begin{document}

\newcommand{\dey}[2]{\frac{\partial #1}{\partial #2}}
\newcommand{\mjscomment}[1]{{\color{blue}*** #1 ***}}
\newcommand{\cgpcomment}[1]{{\color{red}*** #1 ***}}
\newcommand{\mkscomment}[1]{{\color{green}*** #1 ***}}
\newcommand{\PI}{\mbox{$A_0$}}
\newcommand{\SI}{\mbox{$S_0$}}
\newcommand{\BI}{\mbox{$B_0$}}
\newcommand{\PD}{\mbox{$A_{\rm d}$}}

\title{Quasi-solitons and stable superluminal opto-acoustic pulses in Brillouin scattering}

\author{Antoine F. J. Runge$^{1,\dagger}$,*}
\author{Miko\l aj K. Schmidt$^{2}$}
\author{Alexander S. Solntsev$^{1}$}
\author{Michael J. Steel$^{2}$}
\author{Christopher G. Poulton$^{1}$}

\affiliation{
$^{1}$School of Mathematical and Physical Sciences, University of Technology Sydney, NSW 2007, Australia\\
$^{2}$MQ Photonics Research Centre, School of Mathematical and Physical Sciences, Macquarie University, North Ryde NSW 2109, Australia\\
$^{\dagger}$Currently at Institute of Photonics and Optical Science (IPOS), School of Physics, University of Sydney, NSW 2006, Australia\\
$^{*}$Corresponding author: Chris.Poulton@uts.edu.au, antoine.runge@sydney.edu.au
}

\date{\today}

\begin{abstract}
We theoretically and numerically study the evolution of soliton-like waves supported by stimulated Brillouin scattering. First, the emergence and unusual behaviour of resonant solitary waves are investigated for both backward and forward three wave interactions. We find that these waves can be characterized by the ratio between the optical and acoustic damping coefficients. We also examine a second class of non-resonant anti-symmetric soliton-like waves, which
have a more complicated pulse shape than traditional solitons.
These waves are superluminal, with pulse velocities that 
can be tuned by the input Stokes and pump fields. We discuss the excitation of these types of waves and the physical conditions required for their observation.
\end{abstract}

\maketitle



\section{\label{sec:Intro}Introduction}

Stimulated Brillouin scattering (SBS) arises from the coherent coupling between two optical waves and one acoustic wave, and is one of the strongest nonlinear optical processes~\cite{Boyd_NO}. 
SBS plays an important role in
narrow-linewidth laser systems and in many nonlinear fiber optics applications~\cite{Hill_1976, Smith_1991, Okawachi_2005}. Recently, it has attracted renewed interest in the field of integrated photonics~\cite{Pant_2011}, because the small dimensions of nanophotonic waveguides allow for a strong coupling between optical and acoustic waves, greatly enhancing the SBS effect~\cite{Eggleton_2019}. Research in integrated SBS photonics has undergone a period of explosive growth in the past few years, with the demonstration of novel miniaturized optical devices~\cite{Otterstrom_2018, Kittlaus_2018, Eggleton_2019, Casas_Bedoya_2015} and advanced applications~\cite{Eggleton_2019, Merklein_2017, Marpaung_2019}. However, despite these advancements, the fundamental understanding of nonlinear multi-wave dynamics supported by the SBS interaction remains incomplete. In particular, here we report a number of novel properties of two classes of solitary wave solutions that can exist in systems exhibiting SBS. 

Brillouin interactions exist in both \emph{forward} and \emph{backward} variations, according to whether the lower frequency ``Stokes'' optical field is co- or counter-propagating with the pump optical field. 
Since in most SBS configurations, the opto-acoustic coupling bandwidth is narrow (of order 10~MHz), the interacting fields are either strictly continuous wave excitations or at least long, steady pulses. The system can thus be described by amplitude functions that are essentially constant in time (but not in space)~\cite{Cotter_1983}. 

However, if at least one of the optical fields is a pulse~\cite{Merklein_2017}, this approximation is no longer valid and a more general description is required~\cite{Picholle_1991, Wolff_2015, Wolff_2021}.
The set of coupled-mode equations describing SBS in this regime is similar to the system describing resonant three-wave interactions in a wide range of physical systems, including plasma physics and parametric frequency conversion~\cite{Siegman_1966, Armstrong_1970, Kaup_1979, Picozzi_2001, Picozzi_2002}, and is well known for exhibiting complex dynamics. Of particular interest, these coupled-mode equations have traveling wave solutions known as {\em quasi-solitons}~\cite{Picholle_1991} or solitary waves --- pulses that propagate without changing their shape. These solutions can occur provided the pump and the Stokes waves experience different loss channels while maintaining similar group velocities. This  situation can occur in forward and backward \emph{intermodal} Brillouin scattering~\cite{Wolff_2021}, in which the pump and Stokes fields occupy different spatial modes.  Characteristic of these quasi-soliton solutions is that one of the interacting fields possesses a ``shock-like'' edge and nonzero amplitude at infinity, while the other two fields remain spatially localised~\cite{Kaup_1979}. In backward SBS, these quasi-solitary-wave solutions have been identified as  coherent excitations observed in dissipative fiber lasers~\cite{Picholle_1991, Montes_1997, Montes2_1997}, but the forward scattering regime remains less understood. In addition, the complex dynamics in this system allows for a more general classes of solitary waves which may be important in the dynamics of SBS-based devices and which could lead to novel applications for on-chip optical processing~\cite{Poulton_2012, Wolff_2017,Bookchap2}.
  
In this work, we investigate the dynamics of the coupled wave equations that govern both backward and forward Brillouin scattering. We derive a set of analytic solitary waves for both backward and forward Brillouin scattering, and find that these solutions can be characterised by the ratio of the acoustic and optical loss coefficients. We also investigate a new class of localised waves, which differ from the formal quasi-soliton solutions in that they do not have a closed-form analytic expression. These waves are asymmetric in pulse shape and propagate at velocities that can be tuned by changing the input pump strength and the rise time of the Stokes pulse. We show how these waves emerge under experimental conditions, and give a simple analytic relation for their velocities. We conclude with a discussion on the platform requirements for the observation of these waves.

\section{Governing equations and symmetric solitary wave solutions}
We consider a waveguide oriented along the $z$ axis, with the optical fields propagating along this direction. We adopt the slowly-varying envelope approximation for the interacting waves and neglect optical dispersion, due to the extremely narrow linewidth associated with the nanosecond timescales of opto-elastic interactions. Indeed, taking a nominal dispersion for a nanophotonic waveguide of order $|D|=1000$~ps/nm/km and a typical pulse width of $\tau =$100~ns at wavelength $\lambda = 1550$~nm, we find a dispersion length of $L_D = \tau^2/|\beta_2|=2\pi c \tau^2 /(\lambda^2 |D|)=7.8$~km~\cite{Lamont_2008, Judge_2010, Bres_2023}. This is comfortably longer than any chip-based waveguide system we might consider in the medium future, and certainly many orders of magnitude larger than the pulse width.

After suitable scaling (see Appendix~A), the coupled mode equations describing the evolution of the field envelopes can then  be written as~\cite{Wolff_2015,Sipe_2016,Wolff_2021}
\begin{subequations} \label{eq:CMEs}
\begin{align}
\left( \dey{}T + V_1 \dey{}Z + \mu_1 \right) a_1  &= -\kappa \, a_2 b\\
\left( \dey{}T + V_2 \dey{}Z + \mu_2 \right) a_2  &= \kappa \, a_1 b^* \\
\left( \dey{}T + V_3 \dey{}Z + \mu_3 \right) b  &= \kappa \, a_1 a_2^*, \label{eq:normcmec}
\end{align}
\end{subequations}
where $a_1(Z,T)$, $a_2(Z,T)$ and $b(Z,T)$ are the envelopes of the high frequency optical field (pump), low frequency optical field (Stokes), and acoustic field, respectively. Here $V_j$, with $j = {1,2,3}$, are the normalised group velocities associated with the pump, Stokes, and acoustic waves, and $\mu_j$ are the amplitude attenuation coefficients. The constant $\kappa$ is the dimensionless Brillouin coupling coefficient that arises from the overlap of the acoustic and optical fields~\cite{Wolff_2015,Sipe_2016}.
In Eqs.~(\ref{eq:CMEs}a-c) the optical and acoustic powers, time and space dimensions, and coupling and loss
coefficients have all being normalised to natural units. In particular, the dimensionless velocities $V_i$ are expressed in units of the physical group velocity of the pump. The pump velocity $V_1$ is thus always unity, and the sign of $V_2$ determines whether the interaction is forward ($V_2>0$) or backward ($V_2<0$). 
In fact, due to the small frequency shift between the optical fields, we invariably choose the fields to have velocities of the same magnitude $|V_2|=V_1=1$. The normalization procedure leading to these equations is detailed in Appendix~A. The power levels required for realistic experiments  are discussed in Section \ref{sec:Design}.


\subsection{Solitary wave solutions}
We begin with the traveling wave solutions to equations
(\ref{eq:CMEs}a-c). These have been studied for general three-wave interactions in~\cite{Kaup_1979} and for backwards SBS in~\cite{Picholle_1991}.
The main simplifying assumption made in seeking closed-form solutions is that the loss in the pump field is negligible ($\mu_1\approx 0$). 
For SBS, this can be justified if one assumes that the pump and Stokes fields are in different optical modes and so experience different losses. In addition, since the acoustic velocity is much smaller than that of light ($V_3 \ll V_1, V_2$),  this term may be dropped in~\eqref{eq:normcmec}.

\begin{figure}
\includegraphics[width=\columnwidth]{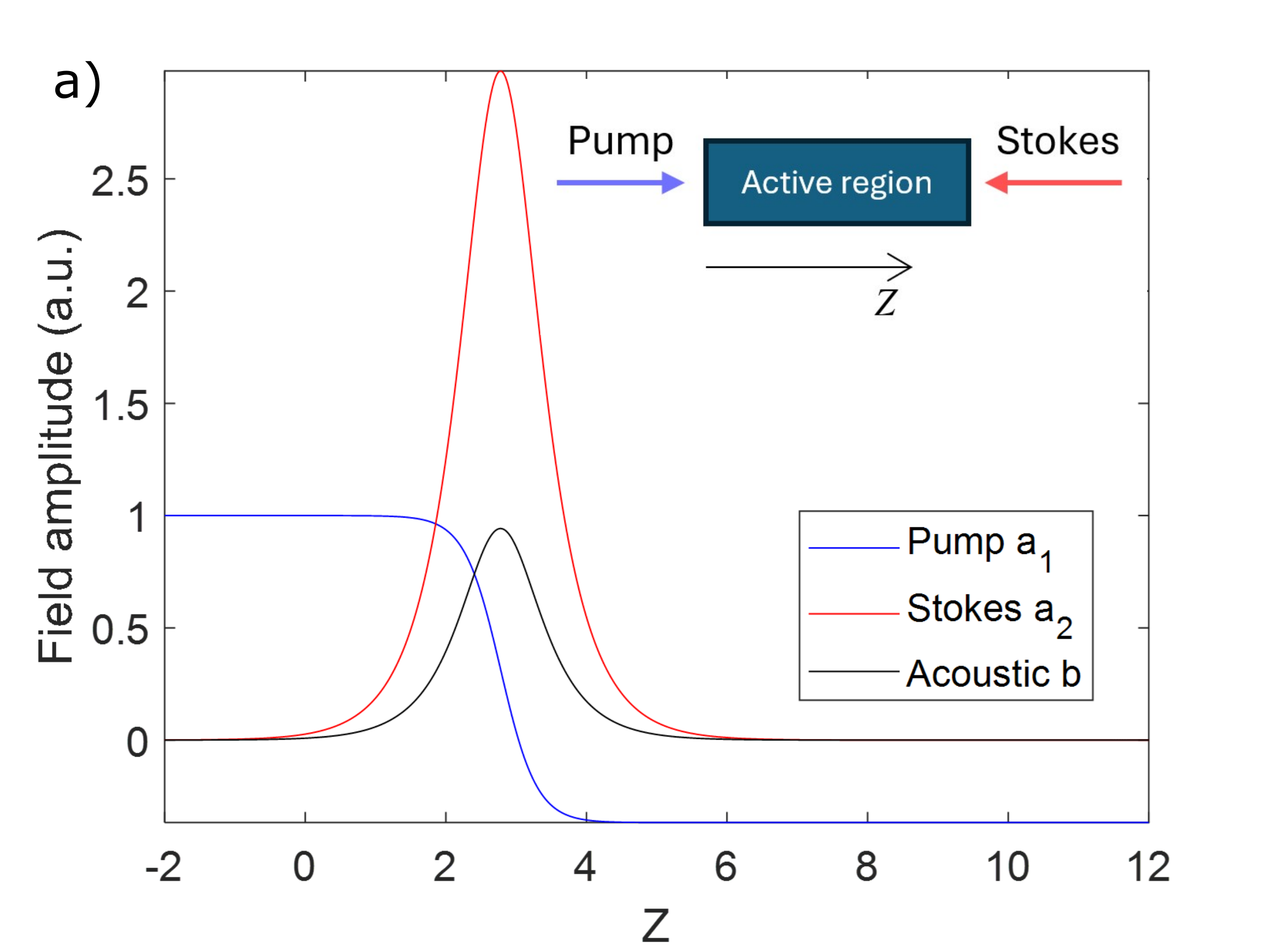}
\includegraphics[width=\columnwidth]{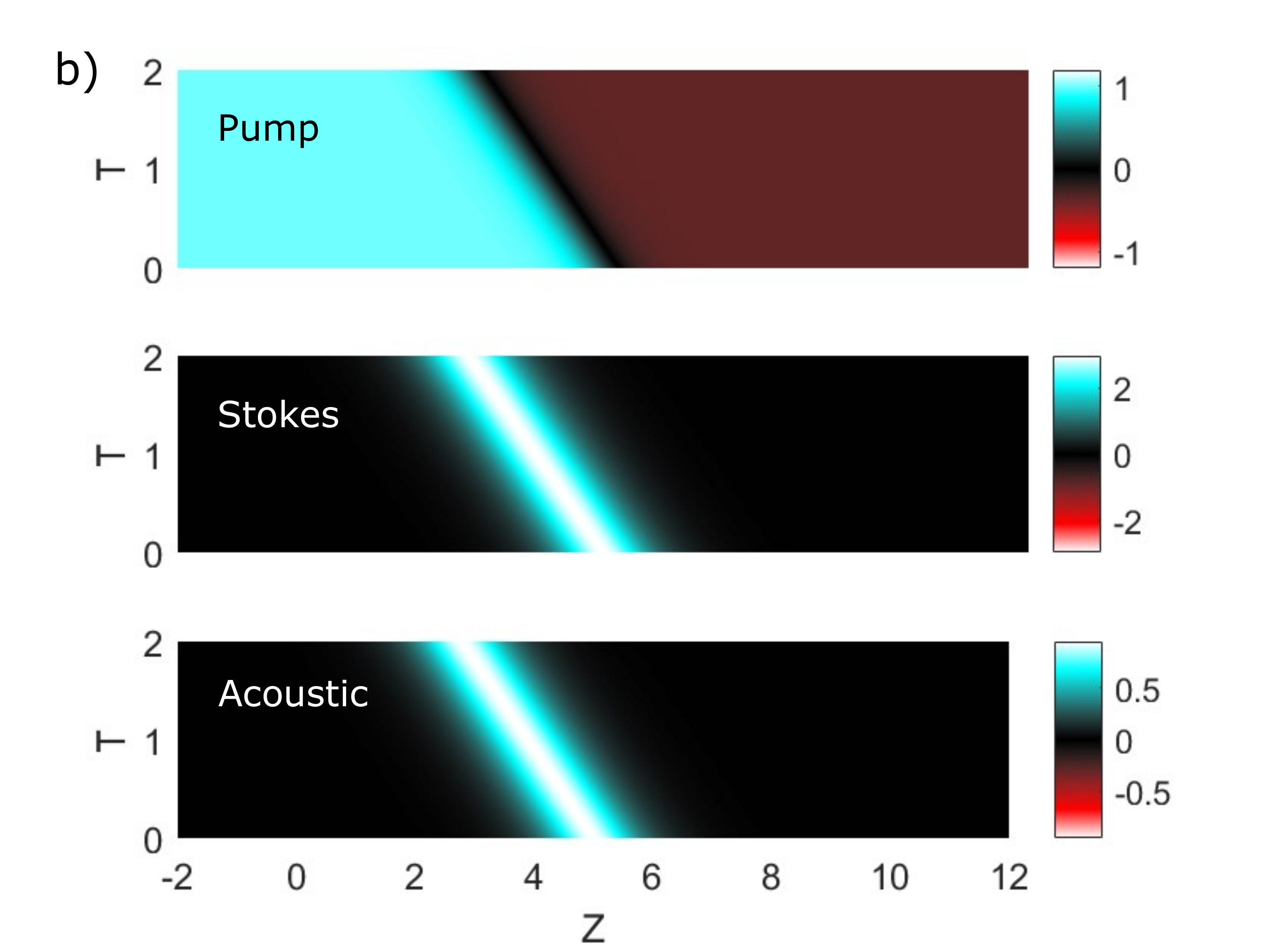}
\caption{\label{fig:steadystate_backwardsQS}
a) Field amplitude profiles for the backward symmetric solitary wave for $\PI = 1.0$, $\mu_2 = 0.1$, and $\mu_3 = 1.0$. The ratio of losses $\mu_2/\mu_3 = 0.1$ leads to a predicted pulse velocity of $V = -1.11$.  A schematic of the pump and Stokes field directions is shown in  inset. b) Field traces demonstrate the stability of the collective excitation and
the pulse velocity $V = -1.11$.}
\end{figure}

\begin{figure}
\includegraphics[width=\columnwidth]{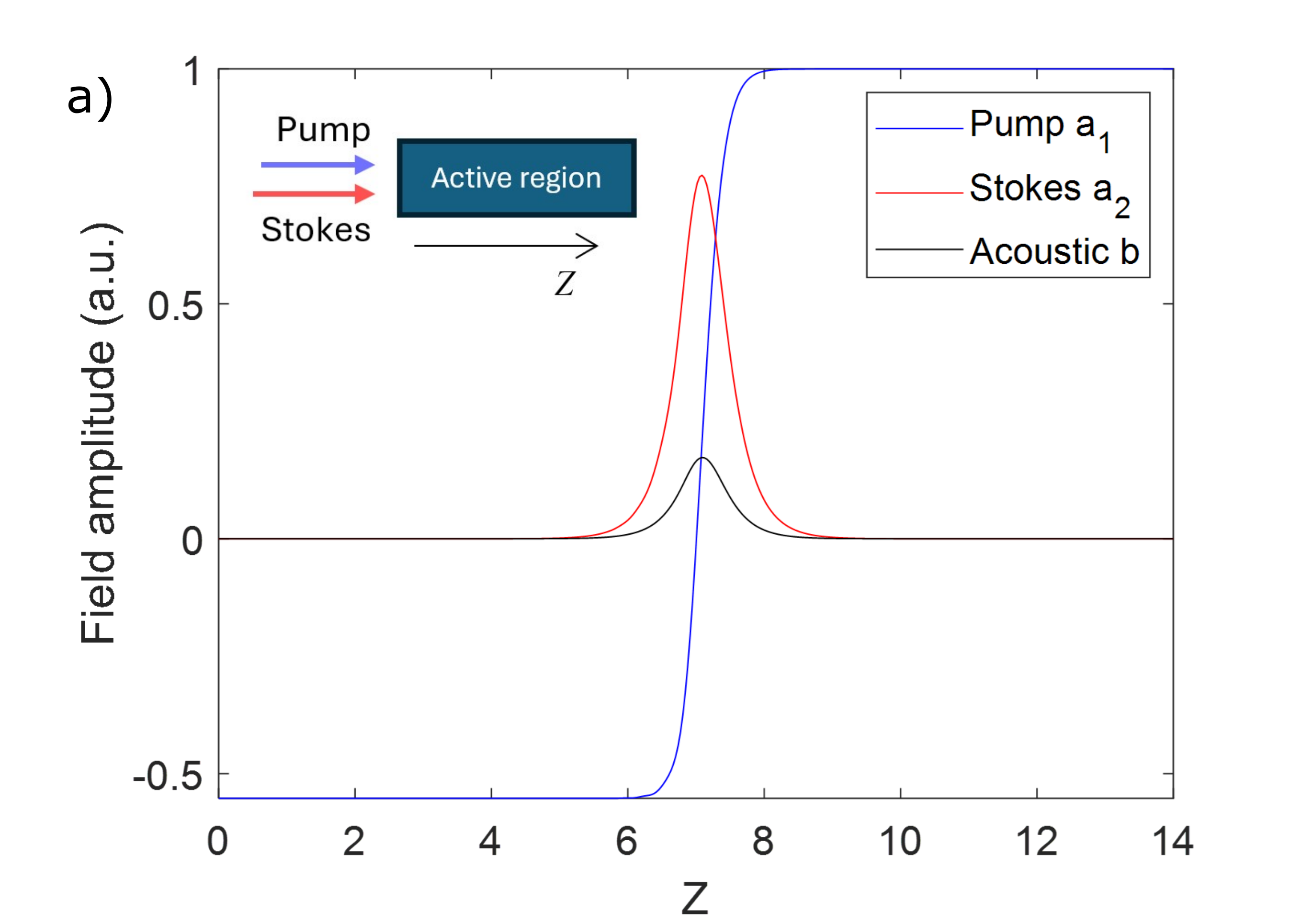}
\includegraphics[width=\columnwidth]{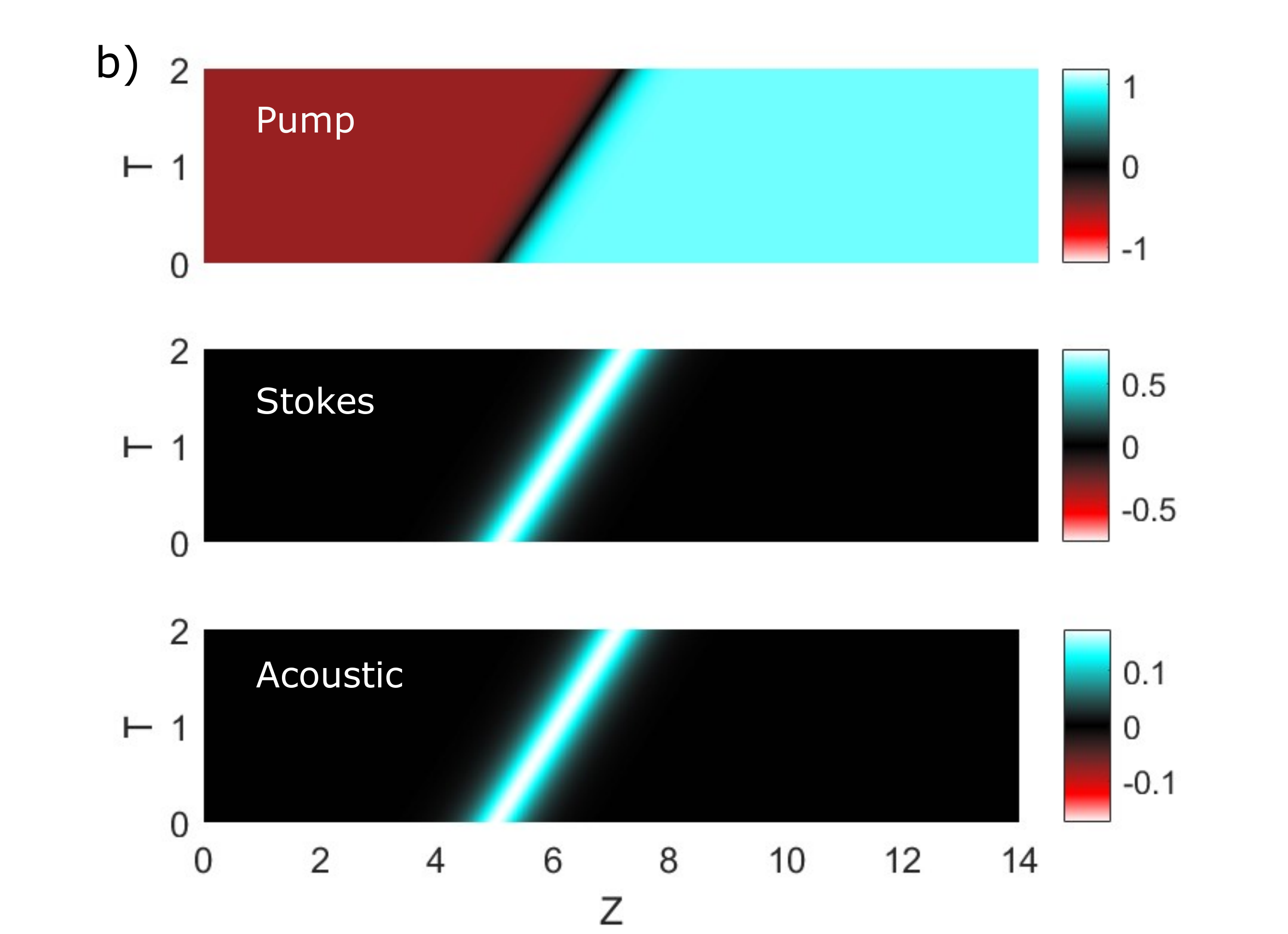}
\caption{\label{fig:steadystate_forwardsQS}
a) Field amplitude profiles for the forward symmetric solitary wave for $\PI = 1.0$, $\mu_2 = 0.2$, and $\mu_3 = 1.0$. The ratio of losses $\mu_2/\mu_3 = 0.2$ leads to a predicted pulse velocity of $V = 1.25$. 
 Note the limiting field from $\PI =1$ occurs as $Z \rightarrow \infty$. A schematic of the pump and Stokes field directions is shown in  inset. b) Field traces demonstrate stability of the combined excitation group and
the  pulse velocity $V = 1.25$.}
\end{figure}

Following~\cite{Picholle_1991}, we look for traveling solutions to Eqs.~(\ref{eq:CMEs}) of the form
\begin{subequations} \label{eq:ansatz1}
\begin{align}
a_1(Z,T) &= \PI - \PD\left[1+ \tanh\left(\frac{Z-VT}{\Delta}\right)\right]
\\
a_2(Z,T) &= S \sech\left(\frac{Z-VT}{\Delta}\right)
\\
b(Z,T) &= B \sech\left(\frac{Z-VT}{\Delta}\right),
\end{align}
\end{subequations}
where $V$ is the velocity of the co-moving disturbance ---
pulse-like for the Stokes and acoustic waves and shock-like for the pump --- and $\Delta$ determines the pulse width. Such a set of solutions requires a continuously injected pump wave for which the constant $\PI$ determines the incoming $a_1(Z \rightarrow - \infty,t) $ or outgoing amplitude  $a_1(Z \rightarrow  \infty,t) $, according to whether $\Delta$ is positive or negative respectively.
Substituting Eq.~(\ref{eq:ansatz1}) into (\ref{eq:CMEs}), we can find expressions for the parameters $V$, $\Delta$, $S$, $\PD$ and $B$, depending only on the damping and coupling coefficients:
\begin{subequations} \label{eq:qsparams1}
\begin{align}
& V = \frac{V_2}{1-\frac{\mu_2}{\mu_3}}, \label{eq:qsparams1a} \\
&\PD = \PI - \frac{\sqrt{\mu_3\mu_2}}{|\kappa|}~,~~\Delta = -\frac{V}{\PD^*}\frac{1}{|\kappa|}\sqrt{\frac{\mu_2}{\mu_3}} \\
& |S|^2 = |\PD|^2 \left(\frac{\mu_3}{\mu_2} - \frac{V_1}{V_2}\frac{\mu_3}{\mu_2} + \frac{V_1}{V_2}\right) ~,~~
B = -\kappa \PD S^*\frac{\Delta}{V} .
\end{align}
\end{subequations}
These five parameters can be successively evaluated given the losses $\mu_2$ and $\mu_3$, the coupling coefficient $\kappa$, and the pump amplitude $\PI$. If we further assume that the
input field $\PI$ is real, it follows that $\PD$ is also real. Equation (\ref{eq:qsparams1}c) contains an arbitrary phase term for $S$ which also determines the phase of $B$. Assuming that this phase is  zero leads to all the fields being real, and the complex conjugates can then be ignored in Eq.~(\ref{eq:qsparams1}).

As will become important, we note from Eq.~\eqref{eq:qsparams1a} that the direction of motion of the excitation indicated by the sign of $V$ is not solely determined by the sign of $V_2$ (ie. by whether the system operates in a forward or backward Brillouin scattering configuration,) but also by the relative losses of the Stokes and acoustic fields.

Figure~\ref{fig:steadystate_backwardsQS} shows the solitary wave 
solution given by Eq.~(\ref{eq:qsparams1}) for a case in the backwards SBS regime ($V_2 = -V_1=-1$). The ratio $\mu_2/\mu_3 = 0.1$ results in
a negative value for the velocity of the excitation velocity $V = -1.11$. Such a solution is sustained by a continuous pump field of amplitude $\PI$ incident on the left side of the waveguide; at the centre of the wave-packet energy is transferred  from the pump to the Stokes, simultaneously exciting an acoustic field. The dissipation in  the Stokes and acoustic modes exactly balances the energy transferred to these fields as the pulse propagates. The pulse shape is retained therefore not via a balance of dispersion and nonlinear effects, as occurs in traditional solitons, but by a balance of energy transfer from the pump and dissipation
of the Stokes and acoustic fields.

Figure~\ref{fig:steadystate_forwardsQS} shows the solitary wave 
solution given by \ref{eq:qsparams1}) for forward SBS
($V_2$ = $V_1$). This time, a ratio $\mu_2/\mu_3 = 0.2$ results in
a positive value for the excitation velocity $V = 1.25$. We observe some qualitatively different features in the forward case via changes in the parameters $\Delta$, $S$ and $\PD$ in (\ref{eq:qsparams1}). First, because $V_1 = V_2$
we immediately see that $S = |\PD|$ i.e. the amplitude of the Stokes is exactly half the rise of the pump. Second, the fact that $V$ is positive means that the width parameter
$\Delta$ becomes negative. This flips the direction of the pump field, so that it approaches the boundary value of $\PI$ as $Z \rightarrow \infty$. This makes sense physically, because a forward-propagating wave will be unaffected at distances far in advance of the excitation. As in the backwards case, the forward-propagating solitary wave
is sustained by a balance between transfer of energy from the pump and dissipation from the other two fields.

\begin{figure}
\includegraphics[width=0.9\columnwidth]{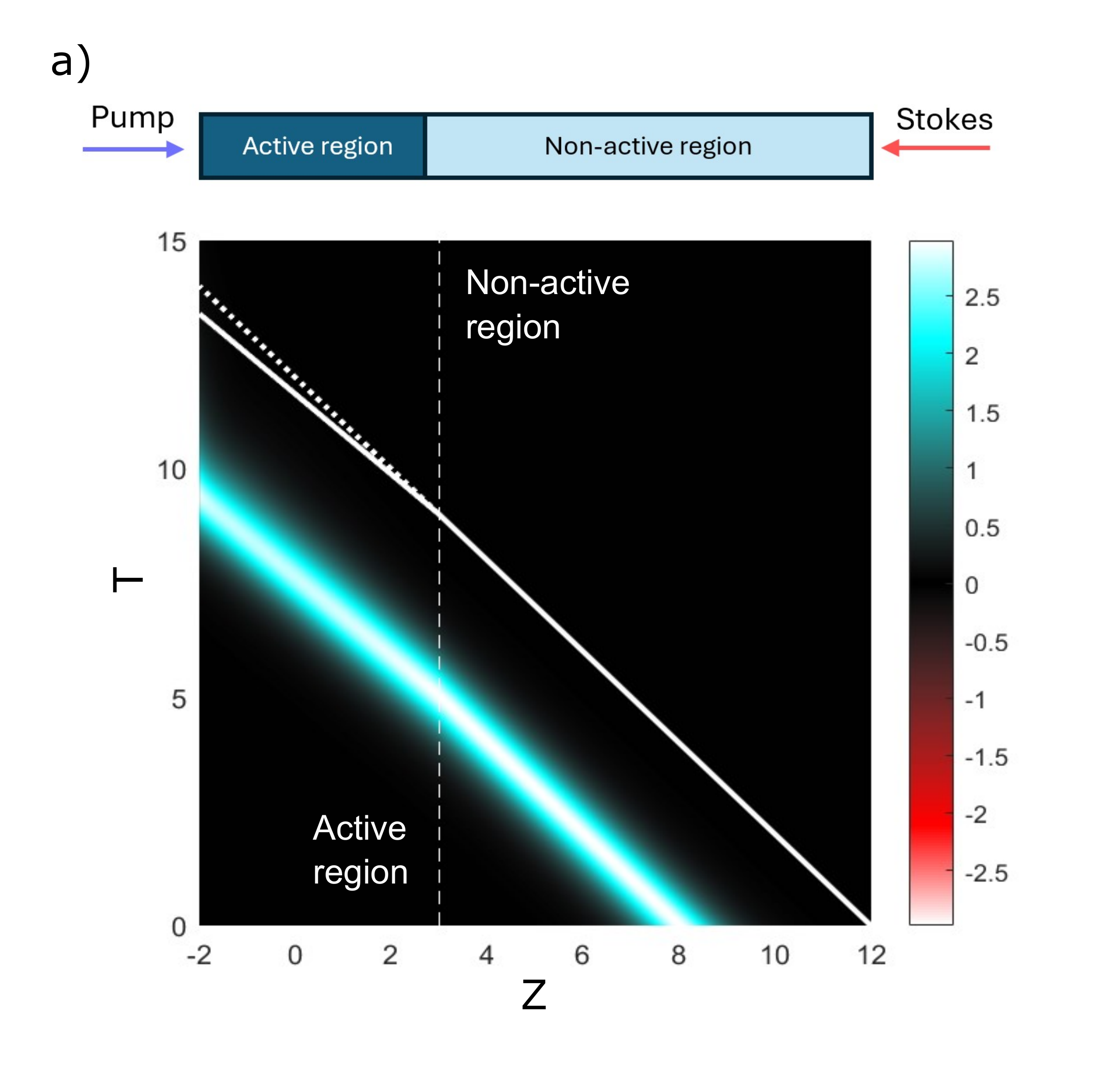}
\includegraphics[width=0.9\columnwidth]{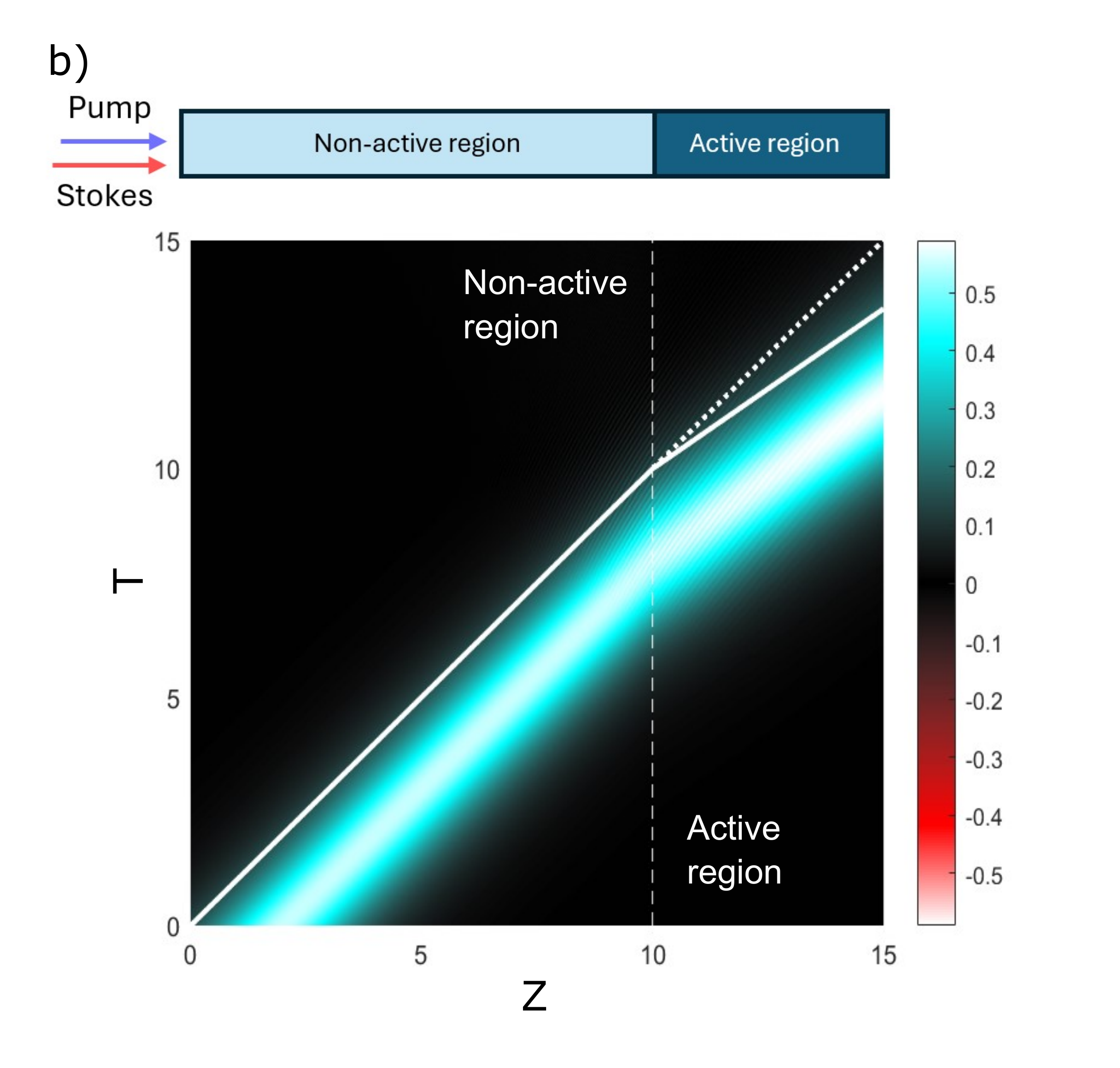}
\caption{\label{fig:externalexcitation}
a) Stokes field of externally excited solitary wave for backward Brillouin scattering. 
Here $\PI = 1.0$, $\mu_2 = 0.1$, and $\mu_3 = 1.0$. 
b) Stokes field of an externally-excited solitary wave for forward Brillouin scattering.
Here $\PI = 1.0$, $\mu_2 = 0.2$, and $\mu_3 = 1.0$. In both (a) and (b) the white line, offset from the pulse peak, shows the increase of the pulse velocity $V$ in the active region (solid line) from the initial velocity $V_2$ in the non-active region (dotted line).}
\end{figure}

In Fig.~\ref{fig:externalexcitation}, we show the formation of solitary waves in the case where both pump and Stokes fields are incident from  outside  the nonlinear region. 
To generate these fields, the initial Stokes field is set to a hyperbolic secant of 
amplitude $S$ and width $\Delta$ determined via Eq.~(\ref{eq:qsparams1}); for backward SBS this leads to 
$S = 2.9805$, $\Delta = 0.5139$, and for forward SBS we find $S = 0.5528$, $\Delta = -1.0113$. For backward SBS the initial pump is set to a CW source of amplitude 
$\PI=1.0$; for the forward case we set the initial pump
to a hyperbolic tangent with a drop off synchronized to the centre of the Stokes pulse, and with $\PI=1.0$ and $P_\mathrm{d} =0.5528$ given in accordance with
Eq.~(\ref{eq:qsparams1}). The initial acoustic field is set to zero. 
In both cases one can see the formation of the superluminal pulse as the acoustic mode builds from the edge of the waveguide. The velocity of the excited pulse can be measured numerically by using a least-squares fit to the position of the peak; the measured velocity in each case is found to be equal to the value of V predicted by (\ref{eq:qsparams1}a) to within the uncertainty of the fit.

\begin{figure}
\includegraphics[width=\columnwidth]{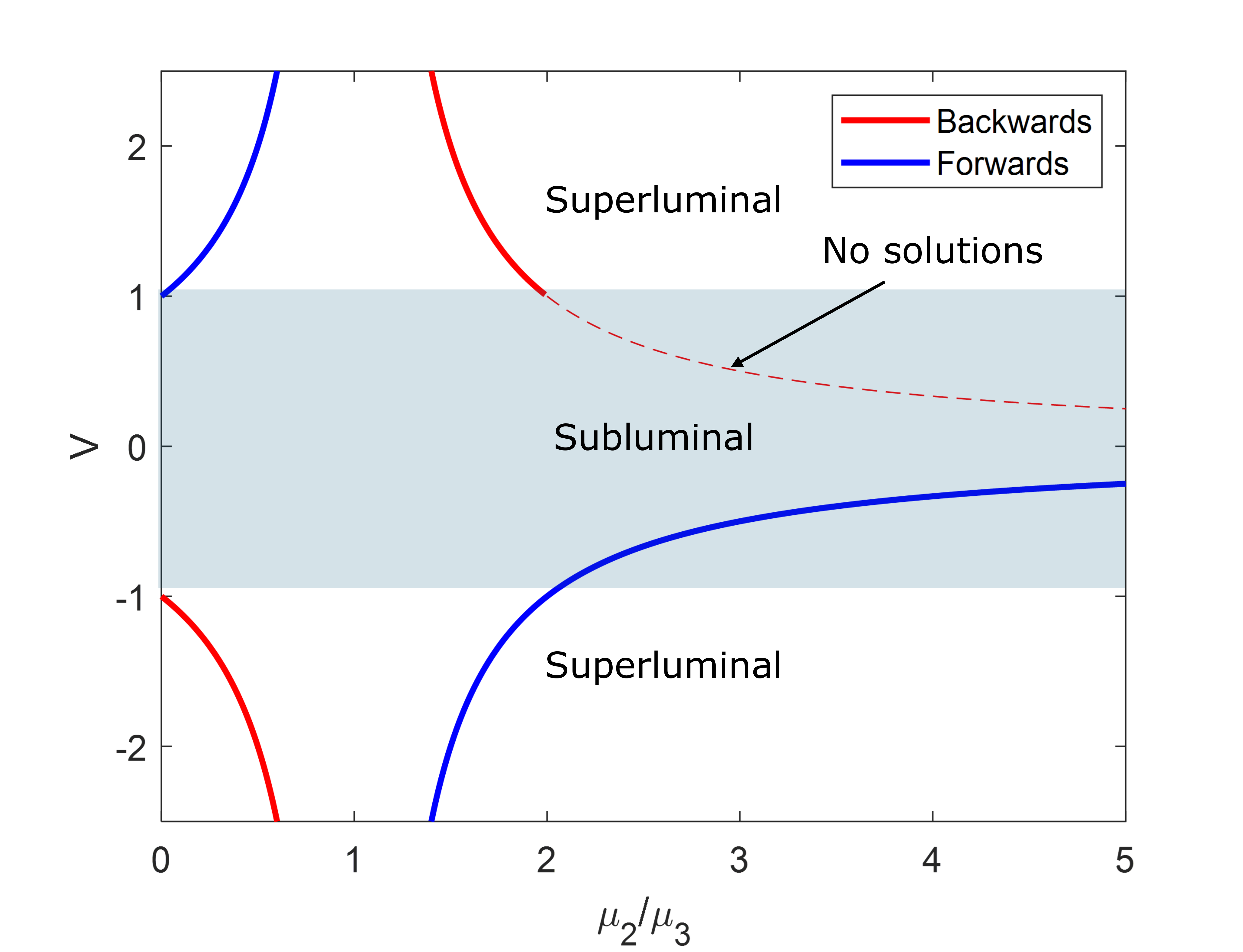}
\caption{
\label{fig:vmap}
Solitary wave pulse velocity $V$ according to Eq.~\eqref{eq:qsparams1a} as a function of the ratio between optical and acoustic attenuation $\mu_2/\mu_3$, for backward ($V_2 = -V_1$) and forward ($V_2 = V_1$) cases. For backward SBS only superluminal solutions are possible, due to the restrictions arising from Eqs (\ref{eq:qsparams1}), whereas for the forward case
subluminal solutions can be found for waveguides with high optical loss ($\mu_2/\mu_3 > 1$).}
\end{figure}

From Eq.~(\ref{eq:qsparams1}a) the velocity $V$ of the solitary wave depends on the velocity of the Stokes field $V_2$ as well as on the ratio between the optical and acoustic losses $\mu_2/\mu_3$. We show this behavior in Fig. 
\ref{fig:vmap}.
We observe distinct behaviours depending on whether the Brillouin process is forward ($V_2 = V_1$) or backward ($V_2 = -V_1$), and on the material composition of the waveguide. We observe that, by positivity of the term in the brackets in (\ref{eq:qsparams1}d), there are no solutions in the backwards case if $\mu_2/\mu_3>2$.
The stable backward solitary waves are therefore all superluminal. By contrast, the forwards case contains both superluminal $V>1$ and subluminal $V<1$ modes, transitioning smoothly between the two regimes when $\mu_2 = 2\mu_3$. 

\section{Asymmetric solitary wave solutions}
\begin{figure}
\includegraphics[width=\columnwidth]{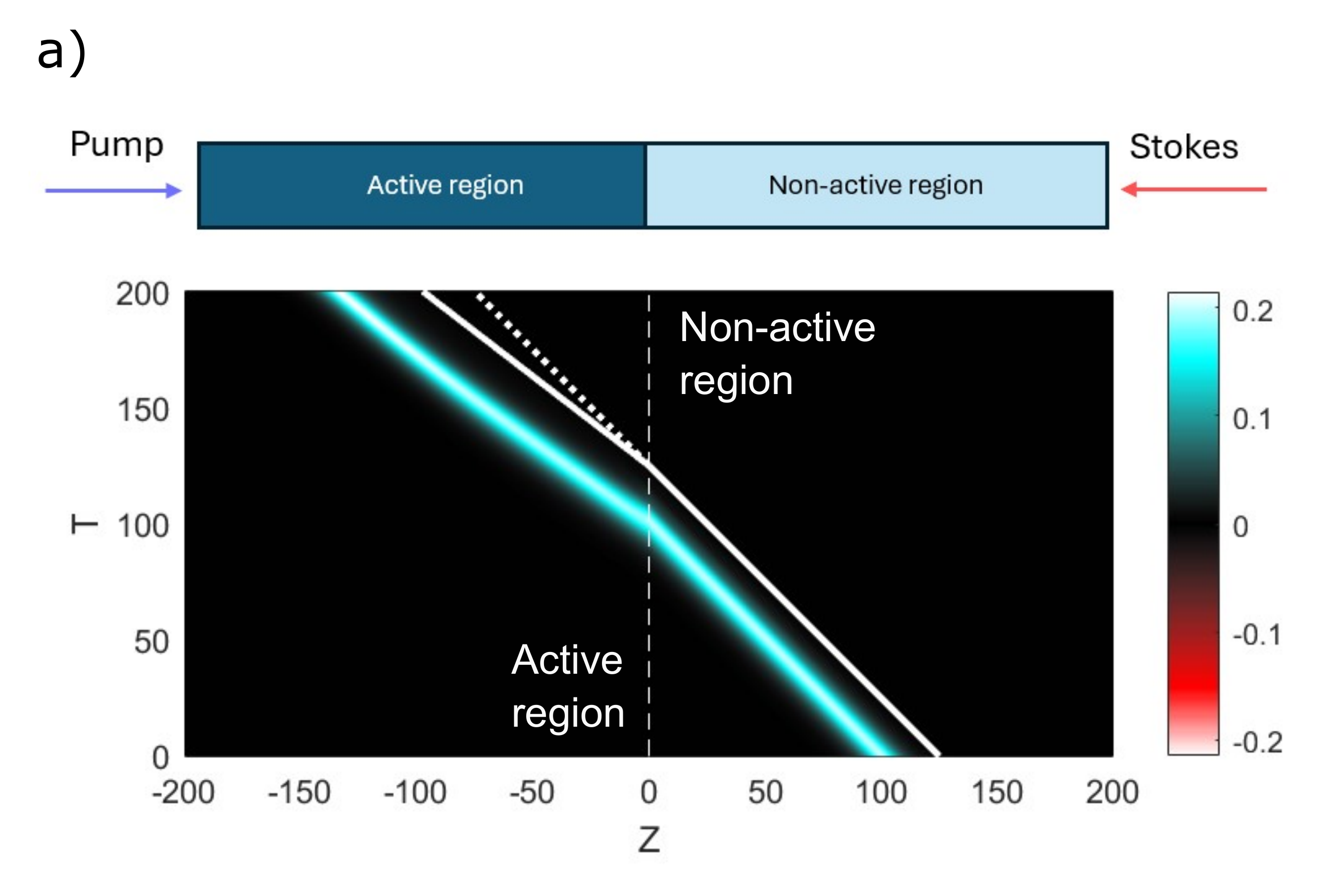}
\includegraphics[width=\columnwidth]{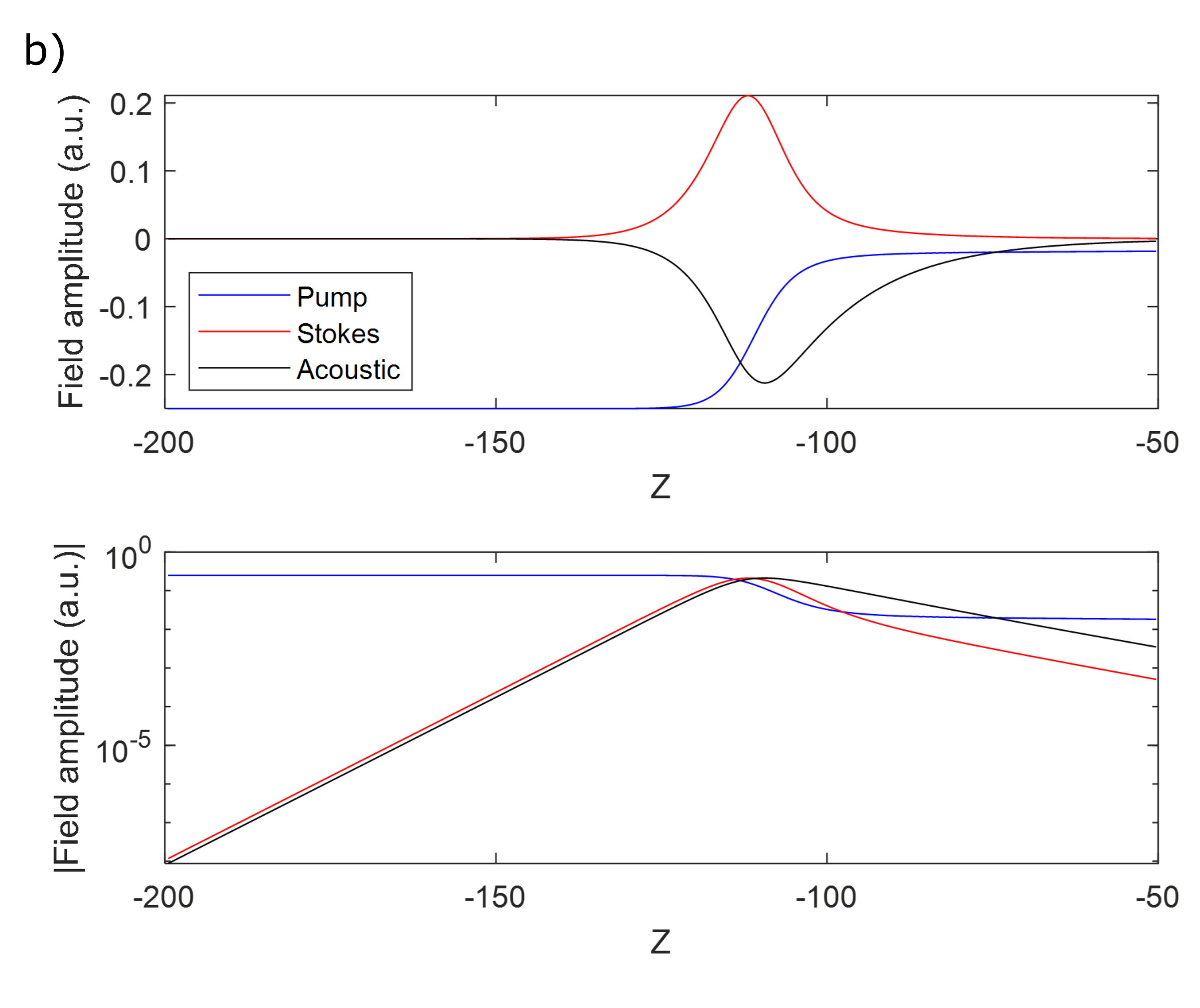}
\caption{\label{fig:asymm_excitation}
a) Stokes field showing the excitation of an asymmetric solitary wave for the backwards SBS case. The material parameters are $\mu_2 = 0.15$, $\mu_3 = 0.1$, and the input pump field is set using $\PI=-0.25$.
Within the active region (left) the fields coalesce about a stable disturbance with velocity $V \approx -1.30$.
b) Field profiles on linear (top) and logarithmic scales, demonstrating the asymmetry between the leading and trailing edges of the wave-packet:
the leading edges of both Stokes and acoustic waves of the
wave-packet increase at the same rate as the incident Stokes pulse $1/\Delta$, whereas the trailing edge of the wave-packet decreases at rate equal to $\mu_3/V$.
}
\end{figure}

We now turn to a class of asymmetric solitary-wave solutions supported by this system that to our knowledge is previously unreported.
In Fig.~\ref{fig:asymm_excitation} we show the numerical solution to the coupled mode equations
(\ref{eq:CMEs}) with a constant pump from the left coinciding with a Stokes pulse from the right hand side. The input Stokes
pulse amplitude is a hyperbolic secant
$a_2(Z,0) = S \sech((Z-Z_0)/\Delta)$, with input Stokes amplitude $S=0.2$, initial position $Z_0 = 100$ and pulse width $\Delta = 5.0$ in normalised units. The attenuation constants are $\mu_2 = 0.15$, $\mu_3 = 0.1$, and the input pump field  from the left is set to a level of $\PI=-0.25$. 
It can be observed that a superluminal wave-packet is formed with a specific velocity $V \approx -1.30$, which is different to that expected for the symmetric solitary wave studied in the previous section. In fact there are no negative velocity symmetric analytic solutions for the ratio $\mu_2/\mu_3 = 1.5$.
Examination of the evolution of this pulse demonstrates that it is stable,
converging to within numerical precision of the final pulse shape in less than 25 normalized time units.

The spatial dependence of this pulse is asymmetric:
looking at the acoustic pulse on a logarithmic scale (Fig. \ref{fig:asymm_excitation}b, lower) we see that the leading edge of the pulse (towards negative $z$) is  straight line with the same
slope in $z$ as that of the incident Stokes wave (equal to $\log_{10} e/\Delta$ on the logarithmic scale, while the trailing edge of the pulse is considerably shallower, and has slope equal to $\log_{10} e\mu_1/V$. 
As with the symmetric solitary waves, the pulse 
demonstrates the characteristics of a shock-wave moving at superluminal velocity through the material: energy is transferred from the pump to the Stokes and to the acoustic wave on the leading edge of the pulse. Unlike the
symmetric case, the trailing edge (towards positive $z$) is dominated by the acoustic tail, which decays at exactly the rate required to
retain the pulse shape for the specific wave-packet velocity $V$
\footnote{Because the moving disturbance ``pulls'' the acoustic field along with it, 
asymmetric disturbances of this type might be designated ``pull-tons''.
Any similarity to the name of one of the authors is purely coincidental.}.

\begin{figure}
\includegraphics[width=\columnwidth]{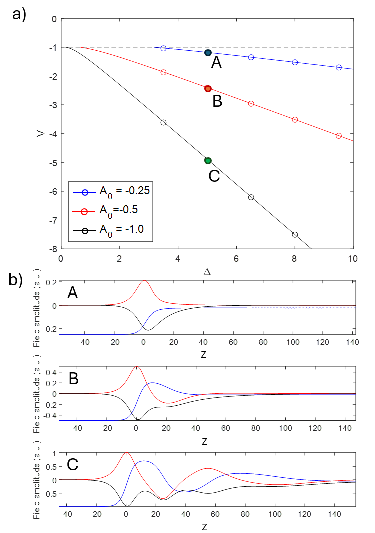}
\caption{\label{fig:deltaV}
a) Pulse velocities, both measured (circles) and according to Eq.~\ref{eq:vequation}) (solid line) for asymmetric pulses as a function of input Stokes width $\Delta$,
plotted for three different input pump powers. Stable solutions exist only  below $V = -1$ (dashed line).
The material parameters are $\mu_2 = 0.15$, $\mu_3 = 0.1$.
b) Pulse fields at the points A, B and C. As the input pump
power increases the pulse becomes more complicated in structure, broadening and increasing in number of nodes in the optical fields. The acoustic mode does not have a node for this class of pulses.
}
\end{figure}

In Fig. \ref{fig:deltaV} (circles) we plot the measured velocities of these asymmetric 
pulses as a function of the Stokes width $\Delta$ and for increasing input pump amplitude $\PI$. 
We see that for higher pump amplitudes $\PI$ and larger values of $\Delta$ the pulses become increasingly superluminal.
The fields at selected points are plotted in Fig. \ref{fig:deltaV}b. We observe that for constant values of $\PI$ the field shape is retained, however as $\PI$ increases there is a smooth transition in the fields to a more complicated structure with a large number of nodes. We also observe that the acoustic mode does not  have a node for this class of pulses.

Although these asymmetric pulses do not appear to have 
analytic solutions, we can gain insight into their behaviour as well as
expressions for their pulse velocities by examining the behaviour of the 
fields at the leading edge of the pulse. To this end, 
we first transform to a reference frame that is co-moving with the
pulse at velocity $V$, i.e. $a_1(Z,T) = a_1(\xi)$, 
$a_2(Z,T) = a_2(\xi)$, and $b(Z,T) = b(\xi)$, 
where $\xi =(Z-VT)$. 
Substitution into the full system~(\ref{eq:CMEs}) gives,
for the case where $\mu_1 = 0$,
\begin{subequations} \label{eq:asymm1}
\begin{align}
&(V_1 - V) \frac{d a_1}{d\xi}  = -\kappa a_2 b \\
&(V_2 - V) \frac{d a_2}{d\xi}   = \kappa a_1 b^* - \mu_2 a_2\\
&(V_3 - V) \frac{d b}{d\xi} = \kappa a_1 a_2^* - \mu_3 b.
\end{align}
\end{subequations}
 Once the velocity $V$ is known, Eqs.~\eqref{eq:asymm1} can be directly integrated to find the pump, Stokes and acoustic fields of the pulse as functions of $\xi$.

To compute the velocity of the pulse, we look for solutions for which the pump is constant, while the Stokes and acoustic field decay exponentially for negative $\xi$, with decay rate given by $\Delta$, as exhibited by the form of the fields in Fig. \ref{fig:asymm_excitation}b. That is, as $\xi \rightarrow -\infty$,
\begin{equation}\label{eq:asympt}
a_1 \rightarrow \PI ~,~~ a_2 \rightarrow \SI e^{\xi/\Delta} ~, b \rightarrow
\BI e^{\xi/\Delta} ~.
\end{equation}
In fact, inspection of Eqs.~(\ref{eq:asymm1}) reveals that the only non-zero solutions
for which $a_1$ remains constant while $a_2$ and $b$ decay in $z$ must either have the form given in Eq.~(\ref{eq:asympt}), or have $\kappa$, $\mu_2$ or $\mu_3$ equal to zero.
Substituting Eq.~(\ref{eq:asympt}) and taking the limit as $\xi \rightarrow -\infty$, we find that  Eq.~(\ref{eq:asymm1}a) 
is automatically satisfied, while Eq.~(\ref{eq:asymm1}b-c) reduce to 
\begin{subequations}
\begin{align}
&\frac{V_2-V}\Delta = \kappa \PI \frac{\BI}{\SI} - \mu_2 \\
&\frac{V_3-V}\Delta = \kappa \PI \frac{\SI}{\BI} - \mu_3 . 
\end{align}
\end{subequations}
Eliminating $\BI/\SI$, we find an expression for the velocity:
\begin{eqnarray} 
V & = & \frac{1}{2} \bigg\{ \left[V_2 + (\mu_2+\mu_3)\Delta \right] 
 \pm  \Big[\left(V_2 + (\mu_2+\mu_3)\Delta \right)^2 
\nonumber \\
 & - &  4 (V_2+\mu_2 \Delta)(V_3 + \mu_3 \Delta) - |\kappa|^2 |\PI|^2 \Delta^2\Big]^{1/2} \bigg\}.
 \label{eq:vequation}
\end{eqnarray}
This equation defines the values of $V$ that describe co-propagating disturbances satisfying Eqs.~(\ref{eq:asymm1})
and having the form given by Eq.~(\ref{eq:asympt}). However not all such solutions are stable, in the sense that they remain finite for all $\xi$. It can immediately be seen, for example, that there are singularities 
in the equation system Eqs.~(\ref{eq:asymm1}) whenever $V = V_j$
for $j = 1,2,3$. Numerical integration of 
the system Eqs.~(\ref{eq:asymm1})  also 
reveals that solutions for which $|V| < |V_2|$ begin to grow and become unbounded as $\xi \rightarrow \infty$. The physical reason for this
appears to be that, if the wave moves too slowly, energy is transferred from the pump to the Stokes too rapidly to be compensated by the loss in the system, causing the Stokes to grow. To obtain a stable solution we therefore take the negative sign of the square root in Eq.~(\ref{eq:vequation}). We show the resulting values of $V$ in Fig. \ref{fig:deltaV}a (solid lines), compared to the numerically measured velocities, obtained from solving the full system of Eqs.~(\ref{eq:CMEs}). 
We can see that there is almost no difference between the
analytic and numerical velocities.

\begin{figure}
\includegraphics[width=\columnwidth]{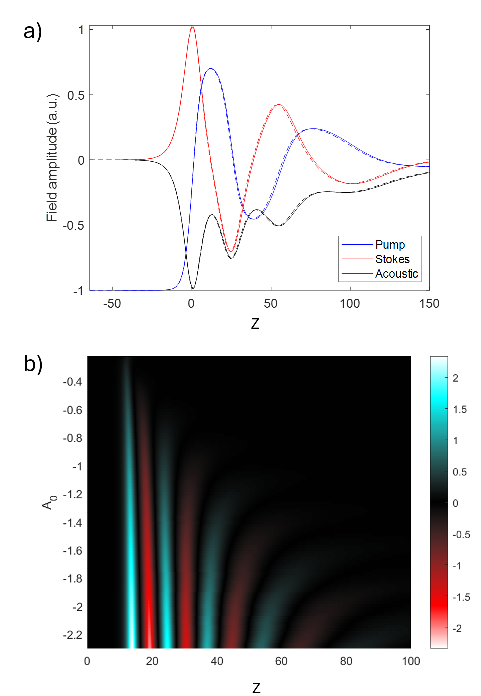}
\caption{\label{fig:pulsecomparison}
a) Comparison of pulse shape obtained from the full solution of the coupled mode 
Eqs.~(\ref{eq:CMEs}) (solid line), with the solution obtained from direct integration of Eqs.~(\ref{eq:asymm1}), using $V$ given by (\ref{eq:vequation}).
The material parameters are $\mu_2 = 0.15$, $\mu_3 = 0.1$.
b) Pulse shape evolution obtained by direct integration of (\ref{eq:asymm1})
for different input pump levels $\PI$. For higher powers the pulse shape extends 
and approaches a periodic structure.
}
\end{figure}

With $V$ determined, we can find approximate values for the fields by integrating Eqs.~(\ref{eq:asymm1}) numerically. We show an example in Fig. \ref{fig:pulsecomparison}a,
for an input pump pulse of $\PI=-1.0$ and Stokes rise time of $\Delta = 5.0$. We compare the computed fields (dotted lines) with the pulse obtained after iterating the full system given in Eqs.(\ref{eq:CMEs}) for a sufficiently long time period that the pulse has stabilized ($T = 125$). Here we note that we have shifted the pulses so that the two maxima coincide. We see that the approximate solution
given by integrating Eqs.~(\ref{eq:asymm1}) agrees very well with the full numerical simulations. Using the computed values of $V$ from Eq.~(\ref{eq:vequation}) together with the 
subsequent numerical integration of Eqs.~(\ref{eq:asymm1}) we can determine the structure of the pulse fields as a function of input pump amplitude $\PI$. This is shown in Fig.~\ref{fig:pulsecomparison}b. We see that for higher powers additional nodes in the Stokes field are created, and the overall length of the pulse becomes longer.

\section{\label{sec:Design}Design challenges}
We discuss now the physical requirements for the excitation and measurement of the solitary waves discussed in the previous sections. We first note (see Appendix) that the field amplitudes have all been normalised to the amplitude of the 
incident pump, so that an amplitude of $A_0=1.0$ corresponds to an input power of 1.0 W. Achieving powers of this order of magnitude is certainly feasible, with typical Brillouin experiments 
having peak powers of around 500~mW for CW or ns pulses~\cite{Pant_2011},
or far higher for shorter pulses~\cite{Stiller2024}.
The power levels discussed throughout this paper are therefore achievable in contemporary Brillouin experiments. 

A more significant challenge is that the optical losses in 
the Stokes must be carefully controlled. Throughout this paper we have assumed, following  Picholle \textit{et al.}~\cite{Picholle_1991},
that the dissipation in the pump and Stokes waves are different. This is in fact a realistic scenario for inter-modal Brillouin scattering, in which 
the losses in different optical modes can differ appreciably~\cite{Madden2007}. The speed $V$ and pulse-length $\Delta$ of the solitary waves depend 
via Eqs.~(\ref{eq:qsparams1}) on the loss ratio $\mu_2/\mu_3$, and 
because the acoustic loss is generally a fixed parameter of Brillouin experiments 
(depending on the phonon lifetime, which is about $\tau_{\rm ph} = 10$~ns
at GHz frequencies), the speed of the quasi-soliton  will depend largely on the dissipation in the Stokes field. To generate a quasi-soliton of speed $V = 2$ 
requires $\mu_2 = \mu_3/2$, which for a waveguide with mode index $n_2 = 2.7$, translates to an optical power loss in the Stokes field of
$\alpha_2 = n_2/4 c \tau_{\rm ph}$, or very close to 1 dB/cm. This is certainly achievable in most Brillouin-active waveguides~\cite{Madden2007} --- the challenge may lie in creating waveguides that are sufficiently lossy in the Stokes field to support the existence of the quasi-soliton.

The final challenge relates to the waveguide length:
the physical pulse length $\Delta_{\rm pulse} = z_0 \Delta $
must be able to fit within the waveguide length, for a duration long enough for the pulse to be excited and to propagate a reasonable distance. For a waveguide with Brillouin gain of $G = 500~$m$^{-1}$W$^{-1}$ (in the typical range for chalcogenide waveguides) we find
(see Appendix) characteristic length scales of $t_0 \approx 850$~ps and $z_0 \approx 9.4$~cm. For an input
pump field of 0.5 W, and pulse velocity $V = 2$ we find from Eqs. (\ref{eq:qsparams1}) that $\Delta_{\rm pulse} \approx$ 64~cm, with a corresponding power in the Stokes of 359~mW. Waveguides of this length are at the very limit of what is currently possible with on-chip structures, but are still within the capabilities of nanostructured optical fibres~\cite{Beugnot2017}. 

\section{\label{sec:Conclusion}Discussion and conclusion}
We have presented two situations where stable pulses arise from solution of the Brillouin coupled mode equations. Both classes of pulses are localized, shape-preserving shock-like disturbances with specific well-defined pulse velocities. The symmetric pulses exist for specific combinations of pump, Stokes and acoustic powers, and so 
qualify as resonant pulses in the same sense as traditional solitons --- these pulses exist as isolated  solutions in the space of all possible combinations of
powers of the three waves. The asymmetric pulses on the other hand are not resonant in this sense: these waves, although localised and stable, will arise for an arbitrary combination of input pump and stokes powers. Interestingly, they can be tuned by the input field to  have an extremely wide range of pulse velocities.

The pulses discussed here have exhibited a variety of superluminal behavior. We note that superluminal motion of the three-wave localized structure does not contradict the special theory of relativity~\cite{thevenaz08,Deng2013,Zhang2011}.
This motion can be viewed as the result of the convective amplification of the leading edges of the Stokes and material pulses, whereas their rears are attenuated, the pump wave being depleted during the interaction and totally or partially restored after that. No transportation of information can be obtained via this deformation process which can only occur if a sufficiently extended background of Stokes light is available.

We expect these results to trigger future studies on solitary waves in SBS. Some of the dynamics described in this work could be observed experimentally in with the current Brillouin-active either in integrated or fiber waveguides~\cite{Madden2007, Beugnot2017}. This could open ways for novel applications for nonlinear all-optical processing~\cite{Eggleton_2019, Marpaung_2019}.


\begin{acknowledgments}
The authors acknowledge funding from the Australian Research Council (ARC) (Discovery
Projects DP160101691, DP200101893, DP220100488, and Discovery Early Career Researcher Awards DE220100509, DE220101272).
\end{acknowledgments}

\appendix

\section{Appendixes}

\subsection{Derivation of normalised coupled mode equations}

Here we derive the normalised form of the Brillouin coupled mode equations in Eqs.~\eqref{eq:CMEs}. We begin with the system of equations appearing in 
\cite{Wolff_2015}:
\begin{subequations} \label{eq:CMEsA1}
\begin{align}
\left( \dey{}t + v_1 \dey{}z + \frac{\alpha_1 v_1}2 \right) A_1  &= - \frac{\omega_1 v_1 Q}{{\cal P}_1}  A_2 B\\
\left( \dey{}t + v_2 \dey{}z + \frac{\alpha_2 v_2}2 \right) A_2  &=  \frac{\omega_2 v_2 Q}{{\cal P}_2} A_1 B^* \\
\left( \dey{}t + v_3 \dey{}z + \frac{\alpha_3 v_3}2 \right) B  &= \frac{\Omega v_3 Q}{{\cal P}_3} A_1 A_2^*,
\end{align}
\end{subequations}
where $(z,t)$ are the physical quantities of distance along the waveguide and time in SI units. Here $v_j$ are the (dimensional) velocities of the three fields and the ${\cal P}_j$ is the chosen normalisation power for each mode (typically chosen as 1~W or 1~mW for each mode). These definitions ensure that the mode amplitudes $A_j$ are dimensionless and the power flow in each field is 
$|A_1|^2{\cal P}_1$, $|A_2|^2{\cal P}_2$, and $|B|^2{\cal P}_3$ respectively. Note that the sign of the power is dependent on the direction of the mode, so that the quantity ${\cal P}_j/v_j$ is always positive. The quantity $Q$ characterises the nonlinear coupling and has units of J/m. Following the definition in~\cite{Wolff_2015} it is in general complex-valued, but it possible to take it real by adding overall constant complex phases to the amplitude functions.

We now normalize the fields to make the coupling term on the right-hand side of Eqs.~(\ref{eq:CMEsA1}) have a common form  in all three equations. This can be done by defining new mode amplitudes
\begin{eqnarray}
a_1(z,t) &=& \sqrt{{\cal P}_1} A_1 ~, ~~~
a_2(z,t) = \sqrt{\frac{\omega_1 v_1  {\cal P}_2}{\omega_2 v_2}} A_2 ~,~~
\nonumber \\
b(z,t) &=& \sqrt{\frac{\omega_1 v_1 {\cal P}_3}{\Omega v_3}} B ~,
\label{adef}
\end{eqnarray}
such that the norm-squared quantities $|a_1|^2$, $|a_2|^2$, $|b|^2$ have dimensions of power, and the second two may easily be scaled to obtain the actual power flowing in the waveguide.

We also scale the space and time coordinates, defining
\begin{equation}
    t = t_0 T ~,~~ z = z_0 Z 
\end{equation}
where $t_0$ and $z_0$ are suitable constants
with dimensions of time and distance respectively. Substituting these,
together with the normalizations given in Eq.~(\ref{adef}), into 
Eqs.~(\ref{eq:CMEsA1}) above gives
\begin{subequations} \label{eq:CMEsA2}
\begin{align}
\left( \dey{}T + \frac{ v_1 t_0}{z_0} \dey{}Z + \frac{\alpha_1 v_1 t_0}2 \right) a_1  = &
- \kappa
a_2 b\\
\left( \dey{}T +  \frac{ v_2 t_0}{z_0} \dey{}Z + \frac{\alpha_2 v_2 t_0}2 \right) a_2  = &\kappa a_1 b^* \\
\left( \dey{}T +  \frac{ v_3 t_0}{z_0} \dey{}Z + \frac{\alpha_3 v_3 t_0}2 \right) b  = &\kappa a_1 a_2^*,
\end{align}
\end{subequations}
where the dimensionless coupling constant is 
\begin{eqnarray}
    \kappa = \frac{Q t_0}{\sqrt{\omega_1 v_1}} \sqrt{\frac{\omega_1 \omega_2 \Omega v_1 v_2 v_3}{{\cal P}_1 {\cal P}_2 {\cal P}_3}}.
\end{eqnarray}
By examining Eqs.~(\ref{eq:CMEsA2}) in the steady state,
assuming the pump is undepleted and that the acoustic mode changes very slowly
with respect to the other fields, we can relate the coupling constant $\kappa$ to the power gain $G$ in the Stokes field in units of m$^{-1}$W$^{-1}$.
The result is
\begin{equation}
\kappa = \frac{t_0}2 \sqrt{v_2 \alpha_3 v_3 G}.
\label{eq:gaineq}
\end{equation}
One can choose $t_0$ to be a typical pulse time, say $t_0 =200$~ps, $z_0 = v_1 t_0$ becomes the distance travelled by the pump in time $t_0$, and $\kappa$ can be evaluated to yield the dimensionless coupling strength.
For gain $G=500$~m$^{-1}$W$^{-1}$, a phonon lifetime $1/\alpha_3 v_3 = 10$~ns and a mode index of $n_2 = c/v_2 = 2.7$ we find $\kappa \approx 0.24$.
Alternatively, we can choose a characteristic time scale such that 
$\kappa = 1$ exactly. For the above parameters we find that $t_0 \approx 850$~ps and $z_0 \approx 9.4$~cm. 


Finally, choosing
\begin{eqnarray}
    V_j = \frac{v_j }{|v_1|} ~, ~~
    \mu_j = \frac{\alpha_j v_j t_0}2
\end{eqnarray}
we obtain the system
\begin{subequations} \label{eq:CMEsA3}
\begin{align}
\left( \dey{}T + V_1 \dey{}Z + \mu_1 \right) a_1  &= 
- \kappa
a_2 b\\
\left( \dey{}T +  V_2 \dey{}Z + \mu_2 \right) a_2  &= \kappa a_1 b^* \\
\left( \dey{}T +  V_3 \dey{}Z +\mu_3 \right) b  &= \kappa a_1 a_2^* ~.
\end{align}
\end{subequations}

\bibliography{SBS_sol_refs}

\providecommand{\noopsort}[1]{}\providecommand{\singleletter}[1]{#1}%
\begin{thebibliography}{36}%
\makeatletter
\providecommand \@ifxundefined [1]{%
 \@ifx{#1\undefined}
}%
\providecommand \@ifnum [1]{%
 \ifnum #1\expandafter \@firstoftwo
 \else \expandafter \@secondoftwo
 \fi
}%
\providecommand \@ifx [1]{%
 \ifx #1\expandafter \@firstoftwo
 \else \expandafter \@secondoftwo
 \fi
}%
\providecommand \natexlab [1]{#1}%
\providecommand \enquote  [1]{``#1''}%
\providecommand \bibnamefont  [1]{#1}%
\providecommand \bibfnamefont [1]{#1}%
\providecommand \citenamefont [1]{#1}%
\providecommand \href@noop [0]{\@secondoftwo}%
\providecommand \href [0]{\begingroup \@sanitize@url \@href}%
\providecommand \@href[1]{\@@startlink{#1}\@@href}%
\providecommand \@@href[1]{\endgroup#1\@@endlink}%
\providecommand \@sanitize@url [0]{\catcode `\\12\catcode `\$12\catcode `\&12\catcode `\#12\catcode `\^12\catcode `\_12\catcode `\%12\relax}%
\providecommand \@@startlink[1]{}%
\providecommand \@@endlink[0]{}%
\providecommand \url  [0]{\begingroup\@sanitize@url \@url }%
\providecommand \@url [1]{\endgroup\@href {#1}{\urlprefix }}%
\providecommand \urlprefix  [0]{URL }%
\providecommand \Eprint [0]{\href }%
\providecommand \doibase [0]{https://doi.org/}%
\providecommand \selectlanguage [0]{\@gobble}%
\providecommand \bibinfo  [0]{\@secondoftwo}%
\providecommand \bibfield  [0]{\@secondoftwo}%
\providecommand \translation [1]{[#1]}%
\providecommand \BibitemOpen [0]{}%
\providecommand \bibitemStop [0]{}%
\providecommand \bibitemNoStop [0]{.\EOS\space}%
\providecommand \EOS [0]{\spacefactor3000\relax}%
\providecommand \BibitemShut  [1]{\csname bibitem#1\endcsname}%
\let\auto@bib@innerbib\@empty
\bibitem [{\citenamefont {Boyd}(1995)}]{Boyd_NO}%
  \BibitemOpen
  \bibfield  {author} {\bibinfo {author} {\bibfnamefont {R.~W.}\ \bibnamefont {Boyd}},\ }\href@noop {} {\emph {\bibinfo {title} {Nonlinear Optics}}}\ (\bibinfo  {publisher} {Elsevier},\ \bibinfo {year} {1995})\BibitemShut {NoStop}%
\bibitem [{\citenamefont {Hill}\ \emph {et~al.}(1976)\citenamefont {Hill}, \citenamefont {Kawasaki},\ and\ \citenamefont {Johnson}}]{Hill_1976}%
  \BibitemOpen
  \bibfield  {author} {\bibinfo {author} {\bibfnamefont {K.~O.}\ \bibnamefont {Hill}}, \bibinfo {author} {\bibfnamefont {B.~S.}\ \bibnamefont {Kawasaki}},\ and\ \bibinfo {author} {\bibfnamefont {D.~C.}\ \bibnamefont {Johnson}},\ }\bibfield  {title} {\bibinfo {title} {cw {Brillouin} laser},\ }\href@noop {} {\bibfield  {journal} {\bibinfo  {journal} {Appl. Phys. Lett.}\ }\textbf {\bibinfo {volume} {28}},\ \bibinfo {pages} {608} (\bibinfo {year} {1976})}\BibitemShut {NoStop}%
\bibitem [{\citenamefont {Smith}\ \emph {et~al.}(1991)\citenamefont {Smith}, \citenamefont {Zarinetchi},\ and\ \citenamefont {Ezekiel}}]{Smith_1991}%
  \BibitemOpen
  \bibfield  {author} {\bibinfo {author} {\bibfnamefont {S.~P.}\ \bibnamefont {Smith}}, \bibinfo {author} {\bibfnamefont {F.}~\bibnamefont {Zarinetchi}},\ and\ \bibinfo {author} {\bibfnamefont {S.}~\bibnamefont {Ezekiel}},\ }\bibfield  {title} {\bibinfo {title} {Narrow-linewidth stimulated {Brillouin} fiber laser and applications},\ }\href@noop {} {\bibfield  {journal} {\bibinfo  {journal} {Opt. Lett.}\ }\textbf {\bibinfo {volume} {16}},\ \bibinfo {pages} {393} (\bibinfo {year} {1991})}\BibitemShut {NoStop}%
\bibitem [{\citenamefont {Okawachi}\ \emph {et~al.}(2005)\citenamefont {Okawachi}, \citenamefont {Bigelow}, \citenamefont {Sharping}, \citenamefont {Zhu}, \citenamefont {Schweinsberg}, \citenamefont {Gauthier}, \citenamefont {Boyd},\ and\ \citenamefont {Gaeta}}]{Okawachi_2005}%
  \BibitemOpen
  \bibfield  {author} {\bibinfo {author} {\bibfnamefont {Y.}~\bibnamefont {Okawachi}}, \bibinfo {author} {\bibfnamefont {M.~S.}\ \bibnamefont {Bigelow}}, \bibinfo {author} {\bibfnamefont {J.~E.}\ \bibnamefont {Sharping}}, \bibinfo {author} {\bibfnamefont {Z.}~\bibnamefont {Zhu}}, \bibinfo {author} {\bibfnamefont {A.}~\bibnamefont {Schweinsberg}}, \bibinfo {author} {\bibfnamefont {D.~J.}\ \bibnamefont {Gauthier}}, \bibinfo {author} {\bibfnamefont {R.~W.}\ \bibnamefont {Boyd}},\ and\ \bibinfo {author} {\bibfnamefont {A.~L.}\ \bibnamefont {Gaeta}},\ }\bibfield  {title} {\bibinfo {title} {Tunable all-optical delays via {Brillouin} slow light in an optical fiber},\ }\href@noop {} {\bibfield  {journal} {\bibinfo  {journal} {Phys. Rev. Lett.}\ }\textbf {\bibinfo {volume} {94}},\ \bibinfo {pages} {153902} (\bibinfo {year} {2005})}\BibitemShut {NoStop}%
\bibitem [{\citenamefont {Pant}\ \emph {et~al.}(2011)\citenamefont {Pant}, \citenamefont {Poulton}, \citenamefont {Choi}, \citenamefont {Mcfarlane}, \citenamefont {Hile}, \citenamefont {Li}, \citenamefont {Thevenaz}, \citenamefont {Luther-Davies}, \citenamefont {Madden},\ and\ \citenamefont {Eggleton}}]{Pant_2011}%
  \BibitemOpen
  \bibfield  {author} {\bibinfo {author} {\bibfnamefont {R.}~\bibnamefont {Pant}}, \bibinfo {author} {\bibfnamefont {C.~G.}\ \bibnamefont {Poulton}}, \bibinfo {author} {\bibfnamefont {D.-Y.}\ \bibnamefont {Choi}}, \bibinfo {author} {\bibfnamefont {H.}~\bibnamefont {Mcfarlane}}, \bibinfo {author} {\bibfnamefont {S.}~\bibnamefont {Hile}}, \bibinfo {author} {\bibfnamefont {E.}~\bibnamefont {Li}}, \bibinfo {author} {\bibfnamefont {L.}~\bibnamefont {Thevenaz}}, \bibinfo {author} {\bibfnamefont {B.}~\bibnamefont {Luther-Davies}}, \bibinfo {author} {\bibfnamefont {S.~J.}\ \bibnamefont {Madden}},\ and\ \bibinfo {author} {\bibfnamefont {B.~J.}\ \bibnamefont {Eggleton}},\ }\bibfield  {title} {\bibinfo {title} {On-chip stimulated {Brillouin} scattering},\ }\href@noop {} {\bibfield  {journal} {\bibinfo  {journal} {Opt. Express}\ }\textbf {\bibinfo {volume} {19}},\ \bibinfo {pages} {8285} (\bibinfo {year} {2011})}\BibitemShut {NoStop}%
\bibitem [{\citenamefont {Eggleton}\ \emph {et~al.}(2019)\citenamefont {Eggleton}, \citenamefont {Poulton}, \citenamefont {Rakich}, \citenamefont {Steel},\ and\ \citenamefont {Bahl}}]{Eggleton_2019}%
  \BibitemOpen
  \bibfield  {author} {\bibinfo {author} {\bibfnamefont {B.~J.}\ \bibnamefont {Eggleton}}, \bibinfo {author} {\bibfnamefont {C.~G.}\ \bibnamefont {Poulton}}, \bibinfo {author} {\bibfnamefont {P.~T.}\ \bibnamefont {Rakich}}, \bibinfo {author} {\bibfnamefont {M.~J.}\ \bibnamefont {Steel}},\ and\ \bibinfo {author} {\bibfnamefont {G.}~\bibnamefont {Bahl}},\ }\bibfield  {title} {\bibinfo {title} {{Brillouin} integrated photonics},\ }\href@noop {} {\bibfield  {journal} {\bibinfo  {journal} {Nat. Photonics}\ }\textbf {\bibinfo {volume} {13}},\ \bibinfo {pages} {664–677} (\bibinfo {year} {2019})}\BibitemShut {NoStop}%
\bibitem [{\citenamefont {Otterstrom}\ \emph {et~al.}(2018)\citenamefont {Otterstrom}, \citenamefont {Behunin}, \citenamefont {Kittlaus}, \citenamefont {Wang},\ and\ \citenamefont {Rakich}}]{Otterstrom_2018}%
  \BibitemOpen
  \bibfield  {author} {\bibinfo {author} {\bibfnamefont {N.~T.}\ \bibnamefont {Otterstrom}}, \bibinfo {author} {\bibfnamefont {R.~O.}\ \bibnamefont {Behunin}}, \bibinfo {author} {\bibfnamefont {E.~A.}\ \bibnamefont {Kittlaus}}, \bibinfo {author} {\bibfnamefont {Z.}~\bibnamefont {Wang}},\ and\ \bibinfo {author} {\bibfnamefont {P.~T.}\ \bibnamefont {Rakich}},\ }\bibfield  {title} {\bibinfo {title} {A silicon {Brillouin} laser},\ }\href@noop {} {\bibfield  {journal} {\bibinfo  {journal} {Science}\ }\textbf {\bibinfo {volume} {360}},\ \bibinfo {pages} {1113} (\bibinfo {year} {2018})}\BibitemShut {NoStop}%
\bibitem [{\citenamefont {Kittlaus}\ \emph {et~al.}(2018)\citenamefont {Kittlaus}, \citenamefont {Otterstrom}, \citenamefont {Kharel}, \citenamefont {Gertler},\ and\ \citenamefont {Rakich}}]{Kittlaus_2018}%
  \BibitemOpen
  \bibfield  {author} {\bibinfo {author} {\bibfnamefont {E.~A.}\ \bibnamefont {Kittlaus}}, \bibinfo {author} {\bibfnamefont {N.~T.}\ \bibnamefont {Otterstrom}}, \bibinfo {author} {\bibfnamefont {P.}~\bibnamefont {Kharel}}, \bibinfo {author} {\bibfnamefont {S.}~\bibnamefont {Gertler}},\ and\ \bibinfo {author} {\bibfnamefont {P.~T.}\ \bibnamefont {Rakich}},\ }\bibfield  {title} {\bibinfo {title} {Non-reciprocal interband {Brillouin} modulation},\ }\href@noop {} {\bibfield  {journal} {\bibinfo  {journal} {Nat. Photonics}\ }\textbf {\bibinfo {volume} {12}},\ \bibinfo {pages} {613–619} (\bibinfo {year} {2018})}\BibitemShut {NoStop}%
\bibitem [{\citenamefont {Casas-Bedoya}\ \emph {et~al.}(2015)\citenamefont {Casas-Bedoya}, \citenamefont {Morrison}, \citenamefont {Pagani}, \citenamefont {Marpaung},\ and\ \citenamefont {Eggleton}}]{Casas_Bedoya_2015}%
  \BibitemOpen
  \bibfield  {author} {\bibinfo {author} {\bibfnamefont {A.}~\bibnamefont {Casas-Bedoya}}, \bibinfo {author} {\bibfnamefont {B.}~\bibnamefont {Morrison}}, \bibinfo {author} {\bibfnamefont {M.}~\bibnamefont {Pagani}}, \bibinfo {author} {\bibfnamefont {D.}~\bibnamefont {Marpaung}},\ and\ \bibinfo {author} {\bibfnamefont {B.~J.}\ \bibnamefont {Eggleton}},\ }\bibfield  {title} {\bibinfo {title} {Tunable narrowband microwave photonic filter created by stimulated {Brillouin} scattering from a silicon nanowire},\ }\href@noop {} {\bibfield  {journal} {\bibinfo  {journal} {Opt. Lett.}\ }\textbf {\bibinfo {volume} {40}},\ \bibinfo {pages} {4154} (\bibinfo {year} {2015})}\BibitemShut {NoStop}%
\bibitem [{\citenamefont {Merklein}\ \emph {et~al.}(2017)\citenamefont {Merklein}, \citenamefont {Stiller}, \citenamefont {Vu}, \citenamefont {Madden},\ and\ \citenamefont {Eggleton}}]{Merklein_2017}%
  \BibitemOpen
  \bibfield  {author} {\bibinfo {author} {\bibfnamefont {M.}~\bibnamefont {Merklein}}, \bibinfo {author} {\bibfnamefont {B.}~\bibnamefont {Stiller}}, \bibinfo {author} {\bibfnamefont {K.}~\bibnamefont {Vu}}, \bibinfo {author} {\bibfnamefont {S.~J.}\ \bibnamefont {Madden}},\ and\ \bibinfo {author} {\bibfnamefont {B.~J.}\ \bibnamefont {Eggleton}},\ }\bibfield  {title} {\bibinfo {title} {A chip-integrated coherent photonic-phononic memory},\ }\href@noop {} {\bibfield  {journal} {\bibinfo  {journal} {Nat. Commun.}\ }\textbf {\bibinfo {volume} {8}},\ \bibinfo {pages} {574} (\bibinfo {year} {2017})}\BibitemShut {NoStop}%
\bibitem [{\citenamefont {Marpaung}\ \emph {et~al.}(2019)\citenamefont {Marpaung}, \citenamefont {Yao},\ and\ \citenamefont {Capmany}}]{Marpaung_2019}%
  \BibitemOpen
  \bibfield  {author} {\bibinfo {author} {\bibfnamefont {D.}~\bibnamefont {Marpaung}}, \bibinfo {author} {\bibfnamefont {J.}~\bibnamefont {Yao}},\ and\ \bibinfo {author} {\bibfnamefont {J.}~\bibnamefont {Capmany}},\ }\bibfield  {title} {\bibinfo {title} {Integrated microwave photonics},\ }\href@noop {} {\bibfield  {journal} {\bibinfo  {journal} {Nat. Photonics}\ }\textbf {\bibinfo {volume} {13}},\ \bibinfo {pages} {80–90} (\bibinfo {year} {2019})}\BibitemShut {NoStop}%
\bibitem [{\citenamefont {Cotter}(1983)}]{Cotter_1983}%
  \BibitemOpen
  \bibfield  {author} {\bibinfo {author} {\bibfnamefont {D.}~\bibnamefont {Cotter}},\ }\bibfield  {title} {\bibinfo {title} {Stimulated \textsc{B}rillouin scattering in monomode optical fiber},\ }\href@noop {} {\bibfield  {journal} {\bibinfo  {journal} {J. Opt. Commun.}\ }\textbf {\bibinfo {volume} {4}},\ \bibinfo {pages} {10} (\bibinfo {year} {1983})}\BibitemShut {NoStop}%
\bibitem [{\citenamefont {Picholle}\ \emph {et~al.}(1991)\citenamefont {Picholle}, \citenamefont {Montes}, \citenamefont {Leycuras}, \citenamefont {Legrand},\ and\ \citenamefont {Botineau}}]{Picholle_1991}%
  \BibitemOpen
  \bibfield  {author} {\bibinfo {author} {\bibfnamefont {E.}~\bibnamefont {Picholle}}, \bibinfo {author} {\bibfnamefont {C.}~\bibnamefont {Montes}}, \bibinfo {author} {\bibfnamefont {C.}~\bibnamefont {Leycuras}}, \bibinfo {author} {\bibfnamefont {O.}~\bibnamefont {Legrand}},\ and\ \bibinfo {author} {\bibfnamefont {J.}~\bibnamefont {Botineau}},\ }\bibfield  {title} {\bibinfo {title} {Observation of dissipative superluminous solitons in a {Brillouin} fiber ring laser},\ }\href@noop {} {\bibfield  {journal} {\bibinfo  {journal} {Phys. Rev. Lett.}\ }\textbf {\bibinfo {volume} {66}},\ \bibinfo {pages} {1454} (\bibinfo {year} {1991})}\BibitemShut {NoStop}%
\bibitem [{\citenamefont {Wolff}\ \emph {et~al.}(2015)\citenamefont {Wolff}, \citenamefont {Steel}, \citenamefont {Eggleton},\ and\ \citenamefont {Poulton}}]{Wolff_2015}%
  \BibitemOpen
  \bibfield  {author} {\bibinfo {author} {\bibfnamefont {C.}~\bibnamefont {Wolff}}, \bibinfo {author} {\bibfnamefont {M.~J.}\ \bibnamefont {Steel}}, \bibinfo {author} {\bibfnamefont {B.~J.}\ \bibnamefont {Eggleton}},\ and\ \bibinfo {author} {\bibfnamefont {C.~G.}\ \bibnamefont {Poulton}},\ }\bibfield  {title} {\bibinfo {title} {Stimulated {Brillouin} scattering in integrated photonic waveguides: Forces, scattering mechanisms, and coupled-mode analysis},\ }\href@noop {} {\bibfield  {journal} {\bibinfo  {journal} {Phys. Rev. A}\ }\textbf {\bibinfo {volume} {92}},\ \bibinfo {pages} {013836} (\bibinfo {year} {2015})}\BibitemShut {NoStop}%
\bibitem [{\citenamefont {Wolff}\ \emph {et~al.}(2021)\citenamefont {Wolff}, \citenamefont {Smith}, \citenamefont {Stiller},\ and\ \citenamefont {Poulton}}]{Wolff_2021}%
  \BibitemOpen
  \bibfield  {author} {\bibinfo {author} {\bibfnamefont {C.}~\bibnamefont {Wolff}}, \bibinfo {author} {\bibfnamefont {M.~J.~A.}\ \bibnamefont {Smith}}, \bibinfo {author} {\bibfnamefont {B.}~\bibnamefont {Stiller}},\ and\ \bibinfo {author} {\bibfnamefont {C.~G.}\ \bibnamefont {Poulton}},\ }\bibfield  {title} {\bibinfo {title} {{Brillouin} scattering—theory and experiment: tutorial},\ }\href@noop {} {\bibfield  {journal} {\bibinfo  {journal} {J. Opt. Soc. Am. B B}\ }\textbf {\bibinfo {volume} {38}},\ \bibinfo {pages} {1243} (\bibinfo {year} {2021})}\BibitemShut {NoStop}%
\bibitem [{\citenamefont {Siegman}(1966)}]{Siegman_1966}%
  \BibitemOpen
  \bibfield  {author} {\bibinfo {author} {\bibfnamefont {A.~E.}\ \bibnamefont {Siegman}},\ }\bibfield  {title} {\bibinfo {title} {Obtaining the equations of motion for parametrically coupled oscillators or waves},\ }\href@noop {} {\bibfield  {journal} {\bibinfo  {journal} {Proc. IEEE}\ }\textbf {\bibinfo {volume} {54}},\ \bibinfo {pages} {756} (\bibinfo {year} {1966})}\BibitemShut {NoStop}%
\bibitem [{\citenamefont {Armstrong}\ \emph {et~al.}(1970)\citenamefont {Armstrong}, \citenamefont {Sudhanshu},\ and\ \citenamefont {Shiren}}]{Armstrong_1970}%
  \BibitemOpen
  \bibfield  {author} {\bibinfo {author} {\bibfnamefont {J.~A.}\ \bibnamefont {Armstrong}}, \bibinfo {author} {\bibfnamefont {S.~J.}\ \bibnamefont {Sudhanshu}},\ and\ \bibinfo {author} {\bibfnamefont {N.~S.}\ \bibnamefont {Shiren}},\ }\bibfield  {title} {\bibinfo {title} {Some effects of group-velocity dispersion on parametric interactions},\ }\href@noop {} {\bibfield  {journal} {\bibinfo  {journal} {J. Quantum Electron.}\ }\textbf {\bibinfo {volume} {6}},\ \bibinfo {pages} {123} (\bibinfo {year} {1970})}\BibitemShut {NoStop}%
\bibitem [{\citenamefont {Kaup}\ \emph {et~al.}(1979)\citenamefont {Kaup}, \citenamefont {Reiman},\ and\ \citenamefont {Bers}}]{Kaup_1979}%
  \BibitemOpen
  \bibfield  {author} {\bibinfo {author} {\bibfnamefont {D.~J.}\ \bibnamefont {Kaup}}, \bibinfo {author} {\bibfnamefont {A.}~\bibnamefont {Reiman}},\ and\ \bibinfo {author} {\bibfnamefont {A.}~\bibnamefont {Bers}},\ }\bibfield  {title} {\bibinfo {title} {Space-time evolution of nonlinear three-wave interactions. i. interaction in a homogeneous medium},\ }\href@noop {} {\bibfield  {journal} {\bibinfo  {journal} {Rev. Mod. Phys.}\ }\textbf {\bibinfo {volume} {51}},\ \bibinfo {pages} {275} (\bibinfo {year} {1979})}\BibitemShut {NoStop}%
\bibitem [{\citenamefont {Picozzi}\ and\ \citenamefont {Haelterman}(2001)}]{Picozzi_2001}%
  \BibitemOpen
  \bibfield  {author} {\bibinfo {author} {\bibfnamefont {A.}~\bibnamefont {Picozzi}}\ and\ \bibinfo {author} {\bibfnamefont {M.}~\bibnamefont {Haelterman}},\ }\bibfield  {title} {\bibinfo {title} {Parametric three-wave soliton generated from incoherent light},\ }\href@noop {} {\bibfield  {journal} {\bibinfo  {journal} {Phys. Rev. Lett.}\ }\textbf {\bibinfo {volume} {86}},\ \bibinfo {pages} {2010} (\bibinfo {year} {2001})}\BibitemShut {NoStop}%
\bibitem [{\citenamefont {Picozzi}\ \emph {et~al.}(2002)\citenamefont {Picozzi}, \citenamefont {Montes},\ and\ \citenamefont {Haelterman}}]{Picozzi_2002}%
  \BibitemOpen
  \bibfield  {author} {\bibinfo {author} {\bibfnamefont {A.}~\bibnamefont {Picozzi}}, \bibinfo {author} {\bibfnamefont {C.}~\bibnamefont {Montes}},\ and\ \bibinfo {author} {\bibfnamefont {M.}~\bibnamefont {Haelterman}},\ }\bibfield  {title} {\bibinfo {title} {Coherence properties of the parametric three-wave interaction driven from an incoherent pump},\ }\href@noop {} {\bibfield  {journal} {\bibinfo  {journal} {Phys. Rev. E}\ }\textbf {\bibinfo {volume} {66}},\ \bibinfo {pages} {056605} (\bibinfo {year} {2002})}\BibitemShut {NoStop}%
\bibitem [{\citenamefont {Montes}\ \emph {et~al.}(1997{\natexlab{a}})\citenamefont {Montes}, \citenamefont {Mikhailov}, \citenamefont {Picozzi},\ and\ \citenamefont {Ginovtar}}]{Montes_1997}%
  \BibitemOpen
  \bibfield  {author} {\bibinfo {author} {\bibfnamefont {C.}~\bibnamefont {Montes}}, \bibinfo {author} {\bibfnamefont {A.}~\bibnamefont {Mikhailov}}, \bibinfo {author} {\bibfnamefont {A.}~\bibnamefont {Picozzi}},\ and\ \bibinfo {author} {\bibfnamefont {F.}~\bibnamefont {Ginovtar}},\ }\bibfield  {title} {\bibinfo {title} {Dissipative three-wave structures in stimulated backscattering. i. a subluminous solitary attractor},\ }\href@noop {} {\bibfield  {journal} {\bibinfo  {journal} {Phys. Rev. E}\ }\textbf {\bibinfo {volume} {55}},\ \bibinfo {pages} {1086} (\bibinfo {year} {1997}{\natexlab{a}})}\BibitemShut {NoStop}%
\bibitem [{\citenamefont {Montes}\ \emph {et~al.}(1997{\natexlab{b}})\citenamefont {Montes}, \citenamefont {Picozzi},\ and\ \citenamefont {Bahloul}}]{Montes2_1997}%
  \BibitemOpen
  \bibfield  {author} {\bibinfo {author} {\bibfnamefont {C.}~\bibnamefont {Montes}}, \bibinfo {author} {\bibfnamefont {A.}~\bibnamefont {Picozzi}},\ and\ \bibinfo {author} {\bibfnamefont {D.}~\bibnamefont {Bahloul}},\ }\bibfield  {title} {\bibinfo {title} {Dissipative three-wave structures in stimulated backscattering. ii. superluminous and subluminous solitons},\ }\href@noop {} {\bibfield  {journal} {\bibinfo  {journal} {Phys. Rev. E}\ }\textbf {\bibinfo {volume} {55}},\ \bibinfo {pages} {1092} (\bibinfo {year} {1997}{\natexlab{b}})}\BibitemShut {NoStop}%
\bibitem [{\citenamefont {Poulton}\ \emph {et~al.}(2012)\citenamefont {Poulton}, \citenamefont {Pant}, \citenamefont {Byrnes}, \citenamefont {Fan}, \citenamefont {Steel},\ and\ \citenamefont {Eggleton}}]{Poulton_2012}%
  \BibitemOpen
  \bibfield  {author} {\bibinfo {author} {\bibfnamefont {C.~G.}\ \bibnamefont {Poulton}}, \bibinfo {author} {\bibfnamefont {R.}~\bibnamefont {Pant}}, \bibinfo {author} {\bibfnamefont {A.}~\bibnamefont {Byrnes}}, \bibinfo {author} {\bibfnamefont {S.}~\bibnamefont {Fan}}, \bibinfo {author} {\bibfnamefont {M.~J.}\ \bibnamefont {Steel}},\ and\ \bibinfo {author} {\bibfnamefont {B.~J.}\ \bibnamefont {Eggleton}},\ }\bibfield  {title} {\bibinfo {title} {Design for broadband on-chip isolator using stimulated {B}rillouin scattering in dispersion-engineered chalcogenide waveguides},\ }\href@noop {} {\bibfield  {journal} {\bibinfo  {journal} {Opt. Express}\ }\textbf {\bibinfo {volume} {20}},\ \bibinfo {pages} {21235} (\bibinfo {year} {2012})}\BibitemShut {NoStop}%
\bibitem [{\citenamefont {Wolff}\ \emph {et~al.}(2017)\citenamefont {Wolff}, \citenamefont {Eggleton}, \citenamefont {Steel},\ and\ \citenamefont {Poulton}}]{Wolff_2017}%
  \BibitemOpen
  \bibfield  {author} {\bibinfo {author} {\bibfnamefont {C.}~\bibnamefont {Wolff}}, \bibinfo {author} {\bibfnamefont {B.~S. B.~J.}\ \bibnamefont {Eggleton}}, \bibinfo {author} {\bibfnamefont {M.~J.}\ \bibnamefont {Steel}},\ and\ \bibinfo {author} {\bibfnamefont {C.~G.}\ \bibnamefont {Poulton}},\ }\bibfield  {title} {\bibinfo {title} {Cascaded forward {B}rillouin scattering to all {S}tokes orders},\ }\href@noop {} {\bibfield  {journal} {\bibinfo  {journal} {New J. Phys.}\ }\textbf {\bibinfo {volume} {19}},\ \bibinfo {pages} {023021} (\bibinfo {year} {2017})}\BibitemShut {NoStop}%
\bibitem [{\citenamefont {Wolff}\ \emph {et~al.}(2022)\citenamefont {Wolff}, \citenamefont {Poulton}, \citenamefont {Steel},\ and\ \citenamefont {Wiederhecker}}]{Bookchap2}%
  \BibitemOpen
  \bibfield  {author} {\bibinfo {author} {\bibfnamefont {C.}~\bibnamefont {Wolff}}, \bibinfo {author} {\bibfnamefont {C.~G.}\ \bibnamefont {Poulton}}, \bibinfo {author} {\bibfnamefont {M.~J.}\ \bibnamefont {Steel}},\ and\ \bibinfo {author} {\bibfnamefont {G.}~\bibnamefont {Wiederhecker}},\ }\bibinfo {title} {{Brillouin} scattering}\ (\bibinfo  {publisher} {Elsevier},\ \bibinfo {address} {Amsterdam},\ \bibinfo {year} {2022})\ Chap.~\bibinfo {chapter} {2}\BibitemShut {NoStop}%
\bibitem [{\citenamefont {Lamont}\ \emph {et~al.}(2008)\citenamefont {Lamont}, \citenamefont {Luther-Davies}, \citenamefont {Choi}, \citenamefont {Madden},\ and\ \citenamefont {Eggleton}}]{Lamont_2008}%
  \BibitemOpen
  \bibfield  {author} {\bibinfo {author} {\bibfnamefont {M.~R.~E.}\ \bibnamefont {Lamont}}, \bibinfo {author} {\bibfnamefont {B.}~\bibnamefont {Luther-Davies}}, \bibinfo {author} {\bibfnamefont {D.-Y.}\ \bibnamefont {Choi}}, \bibinfo {author} {\bibfnamefont {S.}~\bibnamefont {Madden}},\ and\ \bibinfo {author} {\bibfnamefont {B.~J.}\ \bibnamefont {Eggleton}},\ }\bibfield  {title} {\bibinfo {title} {Supercontinuum generation in dispersion engineered highly nonlinear ($\gamma$ = 10~/{W}/m) {A}s$_2${S}$_3$ chalcogenide planar waveguide},\ }\href@noop {} {\bibfield  {journal} {\bibinfo  {journal} {Opt. Express}\ }\textbf {\bibinfo {volume} {16}},\ \bibinfo {pages} {14938} (\bibinfo {year} {2008})}\BibitemShut {NoStop}%
\bibitem [{\citenamefont {Judge}\ \emph {et~al.}(2010)\citenamefont {Judge}, \citenamefont {Dekker}, \citenamefont {Pant}, \citenamefont {de~Sterke},\ and\ \citenamefont {Eggleton}}]{Judge_2010}%
  \BibitemOpen
  \bibfield  {author} {\bibinfo {author} {\bibfnamefont {A.~C.}\ \bibnamefont {Judge}}, \bibinfo {author} {\bibfnamefont {S.~A.}\ \bibnamefont {Dekker}}, \bibinfo {author} {\bibfnamefont {R.}~\bibnamefont {Pant}}, \bibinfo {author} {\bibfnamefont {C.~M.}\ \bibnamefont {de~Sterke}},\ and\ \bibinfo {author} {\bibfnamefont {B.~J.}\ \bibnamefont {Eggleton}},\ }\bibfield  {title} {\bibinfo {title} {Soliton self-frequency shift performance in {A}s$_2${S}$_3$ waveguides},\ }\href@noop {} {\bibfield  {journal} {\bibinfo  {journal} {Opt. Express}\ }\textbf {\bibinfo {volume} {18}},\ \bibinfo {pages} {14960} (\bibinfo {year} {2010})}\BibitemShut {NoStop}%
\bibitem [{\citenamefont {Brès}\ \emph {et~al.}(2023)\citenamefont {Brès}, \citenamefont {Torre}, \citenamefont {Grassani}, \citenamefont {Brasch}, \citenamefont {Grillet},\ and\ \citenamefont {Monat}}]{Bres_2023}%
  \BibitemOpen
  \bibfield  {author} {\bibinfo {author} {\bibfnamefont {C.-S.}\ \bibnamefont {Brès}}, \bibinfo {author} {\bibfnamefont {A.~D.}\ \bibnamefont {Torre}}, \bibinfo {author} {\bibfnamefont {D.}~\bibnamefont {Grassani}}, \bibinfo {author} {\bibfnamefont {V.}~\bibnamefont {Brasch}}, \bibinfo {author} {\bibfnamefont {C.}~\bibnamefont {Grillet}},\ and\ \bibinfo {author} {\bibfnamefont {C.}~\bibnamefont {Monat}},\ }\bibfield  {title} {\bibinfo {title} {Supercontinuum in integrated photonics: generation, applications, challenges, and perspectives},\ }\href@noop {} {\bibfield  {journal} {\bibinfo  {journal} {Nanophotonics}\ }\textbf {\bibinfo {volume} {12}},\ \bibinfo {pages} {1199–1244} (\bibinfo {year} {2023})}\BibitemShut {NoStop}%
\bibitem [{\citenamefont {Sipe}\ and\ \citenamefont {Steel}(2016)}]{Sipe_2016}%
  \BibitemOpen
  \bibfield  {author} {\bibinfo {author} {\bibfnamefont {J.~E.}\ \bibnamefont {Sipe}}\ and\ \bibinfo {author} {\bibfnamefont {M.~J.}\ \bibnamefont {Steel}},\ }\bibfield  {title} {\bibinfo {title} {A {Hamiltonian} treatment of stimulated {{Brillouin}} scattering in nanoscale integrated waveguides},\ }\href {https://doi.org/10.1088/1367-2630/18/4/045004} {\bibfield  {journal} {\bibinfo  {journal} {N. J. Phys.}\ }\textbf {\bibinfo {volume} {18}},\ \bibinfo {pages} {045004} (\bibinfo {year} {2016})}\BibitemShut {NoStop}%
\bibitem [{Note1()}]{Note1}%
  \BibitemOpen
  \bibinfo {note} {Because the moving disturbance ``pulls'' the acoustic field along with it, asymmetric disturbances of this type might be designated ``pull-tons''. Any similarity to the name of one of the authors is purely coincidental.}\BibitemShut {Stop}%
\bibitem [{\citenamefont {Stiller}\ \emph {et~al.}(2024)\citenamefont {Stiller}, \citenamefont {Jaksch}, \citenamefont {Piotrowski}, \citenamefont {Merklein}, \citenamefont {Schmidt}, \citenamefont {Vu}, \citenamefont {Ma}, \citenamefont {Madden}, \citenamefont {Steel}, \citenamefont {Poulton},\ and\ \citenamefont {Eggleton}}]{Stiller2024}%
  \BibitemOpen
  \bibfield  {author} {\bibinfo {author} {\bibfnamefont {B.}~\bibnamefont {Stiller}}, \bibinfo {author} {\bibfnamefont {K.}~\bibnamefont {Jaksch}}, \bibinfo {author} {\bibfnamefont {J.}~\bibnamefont {Piotrowski}}, \bibinfo {author} {\bibfnamefont {M.}~\bibnamefont {Merklein}}, \bibinfo {author} {\bibfnamefont {M.~K.}\ \bibnamefont {Schmidt}}, \bibinfo {author} {\bibfnamefont {K.}~\bibnamefont {Vu}}, \bibinfo {author} {\bibfnamefont {P.}~\bibnamefont {Ma}}, \bibinfo {author} {\bibfnamefont {S.}~\bibnamefont {Madden}}, \bibinfo {author} {\bibfnamefont {M.~J.}\ \bibnamefont {Steel}}, \bibinfo {author} {\bibfnamefont {C.~G.}\ \bibnamefont {Poulton}},\ and\ \bibinfo {author} {\bibfnamefont {B.~J.}\ \bibnamefont {Eggleton}},\ }\bibfield  {title} {\bibinfo {title} {Brillouin light storage for 100 pulse widths},\ }\href {https://doi.org/10.1038/s44310-024-00004-x} {\bibfield  {journal} {\bibinfo  {journal} {npj Nanophotonics}\ }\textbf {\bibinfo {volume} {1}},\ \bibinfo {pages} {5} (\bibinfo {year}
  {2024})}\BibitemShut {NoStop}%
\bibitem [{\citenamefont {Madden}\ \emph {et~al.}(2007)\citenamefont {Madden}, \citenamefont {Choi}, \citenamefont {Bulla}, \citenamefont {Rode}, \citenamefont {Luther-Davies}, \citenamefont {Ta'eed}, \citenamefont {Pelusi},\ and\ \citenamefont {Eggleton}}]{Madden2007}%
  \BibitemOpen
  \bibfield  {author} {\bibinfo {author} {\bibfnamefont {S.~J.}\ \bibnamefont {Madden}}, \bibinfo {author} {\bibfnamefont {D.-Y.}\ \bibnamefont {Choi}}, \bibinfo {author} {\bibfnamefont {D.~A.}\ \bibnamefont {Bulla}}, \bibinfo {author} {\bibfnamefont {A.~V.}\ \bibnamefont {Rode}}, \bibinfo {author} {\bibfnamefont {B.}~\bibnamefont {Luther-Davies}}, \bibinfo {author} {\bibfnamefont {V.~G.}\ \bibnamefont {Ta'eed}}, \bibinfo {author} {\bibfnamefont {M.~D.}\ \bibnamefont {Pelusi}},\ and\ \bibinfo {author} {\bibfnamefont {B.~J.}\ \bibnamefont {Eggleton}},\ }\bibfield  {title} {\bibinfo {title} {Long, low loss etched {A}s$_2${S}$_3$ chalcogenide waveguides for all-optical signal regeneration},\ }\href {http://www.ncbi.nlm.nih.gov/pubmed/19550720} {\bibfield  {journal} {\bibinfo  {journal} {Opt. Express}\ }\textbf {\bibinfo {volume} {15}},\ \bibinfo {pages} {14414} (\bibinfo {year} {2007})}\BibitemShut {NoStop}%
\bibitem [{\citenamefont {Beugnot}\ and\ \citenamefont {Sylvestre}(2017)}]{Beugnot2017}%
  \BibitemOpen
  \bibfield  {author} {\bibinfo {author} {\bibfnamefont {J.-C.}\ \bibnamefont {Beugnot}}\ and\ \bibinfo {author} {\bibfnamefont {T.}~\bibnamefont {Sylvestre}},\ }\href@noop {} {\emph {\bibinfo {title} {Shaping Brillouin Light in Specialty Optical Fibers}}},\ edited by\ \bibinfo {editor} {\bibfnamefont {S.}~\bibnamefont {Boscolo}}\ and\ \bibinfo {editor} {\bibfnamefont {C.}~\bibnamefont {Finot}}\ (\bibinfo  {publisher} {Wiley Press,},\ \bibinfo {year} {2017})\BibitemShut {NoStop}%
\bibitem [{\citenamefont {Th\'{e}venaz}(2008)}]{thevenaz08}%
  \BibitemOpen
  \bibfield  {author} {\bibinfo {author} {\bibfnamefont {L.}~\bibnamefont {Th\'{e}venaz}},\ }\bibfield  {title} {\bibinfo {title} {Slow and fast light in optical fibres},\ }\href@noop {} {\bibfield  {journal} {\bibinfo  {journal} {Nat. Photonics}\ }\textbf {\bibinfo {volume} {2}},\ \bibinfo {pages} {1749} (\bibinfo {year} {2008})}\BibitemShut {NoStop}%
\bibitem [{\citenamefont {Deng}\ \emph {et~al.}(2013)\citenamefont {Deng}, \citenamefont {Gao}, \citenamefont {Liao}, \citenamefont {Duan}, \citenamefont {Cheng}, \citenamefont {Suzuki},\ and\ \citenamefont {Ohishi}}]{Deng2013}%
  \BibitemOpen
  \bibfield  {author} {\bibinfo {author} {\bibfnamefont {D.}~\bibnamefont {Deng}}, \bibinfo {author} {\bibfnamefont {W.}~\bibnamefont {Gao}}, \bibinfo {author} {\bibfnamefont {M.}~\bibnamefont {Liao}}, \bibinfo {author} {\bibfnamefont {Z.}~\bibnamefont {Duan}}, \bibinfo {author} {\bibfnamefont {T.}~\bibnamefont {Cheng}}, \bibinfo {author} {\bibfnamefont {T.}~\bibnamefont {Suzuki}},\ and\ \bibinfo {author} {\bibfnamefont {Y.}~\bibnamefont {Ohishi}},\ }\bibfield  {title} {\bibinfo {title} {{Negative group velocity propagation in a highly nonlinear fiber embedded in a stimulated Brillouin scattering laser ring cavity}},\ }\href@noop {} {\bibfield  {journal} {\bibinfo  {journal} {Appl. Phys. Lett.}\ }\textbf {\bibinfo {volume} {103}},\ \bibinfo {pages} {251110} (\bibinfo {year} {2013})}\BibitemShut {NoStop}%
\bibitem [{\citenamefont {Zhang}\ \emph {et~al.}(2011)\citenamefont {Zhang}, \citenamefont {Zhan}, \citenamefont {Qian}, \citenamefont {Liu}, \citenamefont {Shen}, \citenamefont {Hu},\ and\ \citenamefont {Luo}}]{Zhang2011}%
  \BibitemOpen
  \bibfield  {author} {\bibinfo {author} {\bibfnamefont {L.}~\bibnamefont {Zhang}}, \bibinfo {author} {\bibfnamefont {L.}~\bibnamefont {Zhan}}, \bibinfo {author} {\bibfnamefont {K.}~\bibnamefont {Qian}}, \bibinfo {author} {\bibfnamefont {J.}~\bibnamefont {Liu}}, \bibinfo {author} {\bibfnamefont {Q.}~\bibnamefont {Shen}}, \bibinfo {author} {\bibfnamefont {X.}~\bibnamefont {Hu}},\ and\ \bibinfo {author} {\bibfnamefont {S.}~\bibnamefont {Luo}},\ }\bibfield  {title} {\bibinfo {title} {Superluminal propagation at negative group velocity in optical fibers based on {Brillouin} lasing oscillation},\ }\href@noop {} {\bibfield  {journal} {\bibinfo  {journal} {Phys. Rev. Lett.}\ }\textbf {\bibinfo {volume} {107}},\ \bibinfo {pages} {093903} (\bibinfo {year} {2011})}\BibitemShut {NoStop}%
\end{thebibliography}%

\end{document}